\documentclass{ar2e}
\usepackage{epsfig}

\begin{document}


\newcommand{\prl}[1]{{\it Phys.\ Rev.\ Lett.}\ #1:}
\newcommand{\prd}[1]{{\it Phys.\ Rev.\ D}\ #1:}
\newcommand{\plb}[1]{{\it Phys.\ Lett.\ B}\ #1:}
\newcommand{\npb}[1]{{\it Nucl.\ Phys.\ B}\ #1:}
\newcommand{\epj}[1]{{\it Eur.\ Phys.\ Jour.\ C}\ #1:}
\newcommand{\ijmpa}[1]{{\it Int.\ Jour.\ Mod.\ Phys.\ A}#1:}
\newcommand{\ijp}[1]{{\it Int.\ Jour.\ Phys.}\ #1:}
\newcommand{\jpg}[1]{{\it Jour.\ Phys.\ G}#1:}
\newcommand{\jhep}[1]{{\it JHEP}\ #1:}
\newcommand{\B}{{\cal B}}
\newcommand{\bfB}{{\boldmath $B$}}
\newcommand{\bma}[1]{\boldmath{#1}}
\newcommand{\Bb}{\kern 0.18em\overline{\kern -0.18em {\cal B}}}
\newcommand{\Bbar}{\kern 0.18em\overline{\kern -0.18em B}}
\newcommand{\Bz}{\ensuremath{B^0}}
\newcommand{\Bzb}{\ensuremath{\Bbar^0}}
\def\BzBzb{\ensuremath{\Bz {\kern -0.16em \Bzb}}}
\newcommand{\BB}{\ensuremath{B\Bbar}}
\newcommand{\Kz}{\ensuremath{K^0}}
\newcommand{\Kbar}{\kern 0.2em\overline{\kern -0.2em K}}
\newcommand{\Kzb}{\ensuremath{\Kbar^0}}
\newcommand{\qqbar}{\ensuremath{q\overline q}}
\newcommand{\etapr}{\ensuremath{\eta^{\prime}}}
\newcommand{\CP}{\ensuremath{CP}}
\newcommand{\DG}{\ensuremath{\Delta\Gamma}}
\newcommand{\acp}{\ensuremath{A_{\rm CP}}}
\newcommand{\dt}{\ensuremath{\delta t}}
\newcommand{\fL}{\ensuremath{f_L}}

\renewcommand{\refname}{NUMBERED LITERATURE CITED}


\title{Charmless Hadronic \bfB-Meson Decays}

\markboth{Cheng and Smith}{Charmless Hadronic $B$-Meson Decays}

\author{Hai-Yang Cheng$^a$, James G. Smith$^b$
\affiliation{$^a$Institute of Physics, Academia Sinica, Taipei, Taiwan 115, ROC\\
$^b$Department of Physics, University of Colorado, Boulder, CO 80309-0390, USA \\
}}


\begin{abstract}
We give an overview of the experimental measurements and the theoretical
understanding of the branching fractions and \CP-violating asymmetries of
charmless $B$-meson decays.  Most experimetal results are from the BABAR
and Belle experiments during the past decade.  The global features of
these experimental results are
typically well described by the QCD-motivated theories such as QCD
factorization, pQCD and soft-collinear effective theory. The agreement
between theory and experiment is generally satisfactory, though there
remain some unsolved puzzles that pose a great challenge to both theorists
and experimentalists.
\end{abstract}

\maketitle

\section{INTRODUCTION}
Evidence for the $B$ meson was first seen in 1981 \cite{Bdiscover}. For
the next two decades, the ARGUS experiment operating at DESY in Hamburg,
Germany and the CLEO experiment operating at CESR at Cornell University
studied the properties of these mesons and made many important
discoveries including the discovery of \BzBzb\ mixing.  The
first examples of charmless hadronic $B$ decays were seen by CLEO
\cite{chlsB}.  Starting in 2000, there have been many measurements of
these decays with increasing precision by the BABAR experiment operating at
PEP-II at SLAC in California and the Belle experiment operating at KEK in Japan.
There are now nearly 100 of these decays that have been observed with
a statistical significance of at least four standard deviations.
While the early measurements by CLEO were groundbreaking, the errors
are sufficiently large that they have little weight in the present world
averages.  The CDF experiment at FNAL also has measurements for these
decays.  While they are nearly as precise as BABAR and Belle, there are
only measurements for four decay channels.  Thus in tables throughout
this review, we concentrate on the copious measurements from Belle and
BABAR.

The theoretical study of weak nonleptonic decays of a heavy meson is difficult
and involved due to the interplay of short- and long-distance QCD effects. The
effective weak Hamiltonian at the quark level ${\cal H}=\sum c_i(\mu)O_i(\mu)$
is theoretically well under control, where $O_i$ are  four-quark operators
and $c_i(\mu)$ are the Wilson coefficients which incorporate strong-interaction
effects above the scale $\mu$. However, it is a difficult task to evaluate the
hadronic matrix elements of the local operator $O_i$ reliably due to the
nonperturbative QCD effects involved. A simple and widely employed approach is
based on the valence quark assumption and the vacuum-insertion (or
factorization) approximation in which the hadronic matrix elements of two quark
bilinear operators are saturated by the vacuum intermediate states.

Due to the experimental and theoretical efforts in the past decade, qualitative
understanding of nonleptonic charmless $B$ decays has become possible. Since
the $B$ meson is heavy, it is possible to describe the dynamics of hadronic
decays by theories motivated by QCD rather than by phenomenological models.  A
central aspect of those theories is the factorization theorem which allows us to
disentangle short-distance QCD dynamics from nonperturbative hadronic
effects. In the heavy quark limit, matrix elements can be expressed in terms of
certain nonperturbative input quantities such as light cone distribution
amplitudes and transition form factors. Power corrections beyond the heavy
quark limit generally give the major theoretical uncertainties.

In this article we give an overview of the experimental measurements and the
theoretical understanding of the branching fractions and \CP-violating
asymmetries of charmless $B$-meson decays. We begin with the theoretical and
experimental tools necessary for the study of hadronic $B$ decays. This
is followed with a discussion of
2-body, 3-body, quasi-2-body, and baryonic $B$ decays, and
the status of time-dependent \CP-violation measurements and predictions.
We discuss the puzzles that remain unsolved.

\section{FACTORIZATION AND THEORETICAL TOOLS}
In this section we will introduce various approaches that have been employed
for studying the dynamics of hadronic $B\to M_1 M_2$ decays. In the effective
Hamiltonian approach, the decay amplitude is given by
\begin{eqnarray}
A(B\to M_1M_2)={G_F\over \sqrt{2}}\sum \lambda_i c_i(\mu)\langle M_1M_2|O_i|B\rangle(\mu),
\end{eqnarray}
where $\lambda_i$ are Cabibbo-Kobayashi-Maskawa (CKM) \cite{CKM}
matrix elements, $O_i$ are four-quark operators
and $c_i(\mu)$ are the Wilson coefficients which incorporate strong-interaction
effects above the scale $\mu$. A major theoretical issue is how to evaluate
the matrix elements of the four-quark operators $\langle M_1M_2|O_i|B\rangle$.

\subsection{Naive and Generalized Factorization}

A widely used approximation is the so-called ``naive factorization" or
``vacuum-insertion approximation" under which the matrix element
$\langle M_1M_2|O|B\rangle$  is approximated by
$\langle M_1|J_{1\mu}|0\rangle\langle M_2|J^\mu_2|B\rangle$ or
$\langle M_2|J_{1\mu}|0\rangle\langle M_1|J^\mu_2|B\rangle$
with $J_\mu$ being a bilinear current; that is, the matrix element of a
four-quark operator is expressed as a product of a decay constant and a form
factor. Naive factorization is simple but fails to describe color-suppressed
modes. For example, the predicted ratio of
$\Gamma(D^0\to \bar \Kz\pi^0)/\Gamma(D^0\to K^-\pi^+)\approx 3\times 10^{-4}$
is too small compared with the experimental value of 0.55 \cite{Cheng89}. This
is ascribed to the fact that color-suppressed decays receive sizable
nonfactorizable contributions that have been neglected in naive factorization.
Another issue is that the decay amplitude under naive factorization is not
truly physical because the renormalization scale and scheme dependence of
$c_i(\mu)$ are not compensated by that of the matrix element
$\langle M_1M_2|O_i|B\rangle(\mu)$.
In the improved ``generalized factorization" approach \cite{genFA1,genFA2},
nonfactorizable effects are absorbed into the parameter $N_c^{\rm eff}$, the
effective number of colors. This parameter can be empirically determined from
experiment.  Since these early calculations have been replaced with
improved ones discussed below, we will not discuss these early calculations.

\subsection{Theories of Hadronic \bfB\ Decays}
With the advent of heavy quark effective theory,
nonleptonic $B$ decays can be analyzed systematically within the QCD framework.
There are three popular approaches available in this regard: QCD factorization
(QCDF) \cite{BBNS99}, perturbative QCD (pQCD) \cite{pQCD} and soft-collinear
effective theory (SCET) \cite{SCET}. A detailed discussion of these theories
goes beyond the scope of this review, and
the interested reader is referred to the original literature. Basically,
theories of hadronic $B$ decays are based on the ``factorization theorem"
under which the short-distance contributions to the decay amplitudes can be
separated from the process-independent long-distance parts.\footnote{However,
the charming penguin term advocated in SCET violates this factorization
theorem.} In the QCDF approach, nonfactorizable contributions to the hadronic
matrix elements can be absorbed into the effective parameters $a_i$
\begin{eqnarray}
A(B\to M_1M_2)={G_F\over \sqrt{2}}\sum \lambda_i a_i(M_1M_2)\langle M_1M_2|O_i|B\rangle_{\rm fact},
\end{eqnarray}
where $a_i$ are basically the Wilson coefficients
in conjunction with short-distance nonfactorizable corrections such as vertex,
penguin  corrections and hard spectator interactions, and
$\langle M_1M_2|O_i|B\rangle_{\rm fact}$ is the matrix element evaluated under
the factorization (or vacuum insertion) approximation. For penguin operators
$O_{6,8}$, the current operators $J_{1,2}$ are replaced by the scalar or
pseudoscalar densities.
Since power corrections of order $\Lambda_{\rm QCD}/m_b$ are suppressed in the heavy
quark limit, nonfactorizable corrections to nonleptonic decays are calculable.
In the limits of $m_b\to\infty$ and $\alpha_s\to 0$, naive factorization is
recovered in both QCDF and  pQCD approaches.

Power corrections are often plagued by the end-point divergence that in turn
breaks the factorization theorem. For example, the endpoint divergence occurs
at the twist-3 level for the hard spectator scattering amplitude and at the
twist-2 level for the annihilation amplitude. As a consequence, the estimate
of power corrections is generally model dependent and can only be studied in a
phenomenological way. There is also an endpoint singularity in SCET though this
issue can be resolved after introducing the zero-bin subtraction procedure
\cite{Manohar}. In the pQCD approach, the endpoint singularity is cured
by including the parton's transverse momentum.

\subsection{Diagrammatic Approach}

\begin{figure}[htb!]
\centerline{\psfig{figure=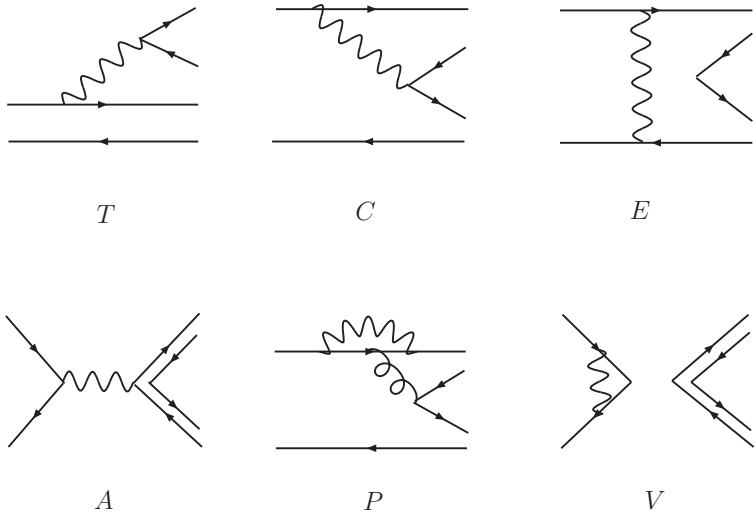,width=4.0in}}
\caption{Various topological diagrams for $B\to M_1M_2$ decays.}
\label{fig:QDS}
\end{figure}

Because a reliable evaluation of hadronic matrix elements is very difficult in
general, an alternative approach is based on the diagrammatic approach. It has
been established that a least
model-dependent analysis of heavy meson decays can be carried
out in the so-called quark-diagram approach \cite{CC86,CC87}. In the
diagrammatic approach, all two-body nonleptonic weak decays
of heavy mesons can be expressed in terms of six distinct quark
diagrams (Fig. \ref{fig:QDS}):\footnote{Historically, the quark-graph
amplitudes $T,\,C,\,E,\,A,\,P$ named in \cite{Gronau94} were originally denoted by
$A,\,B,\,C,\,D,\,E$, respectively, in \cite{CC86,CC87}. For the analysis of charmless $B$ decays, one adds the variants of the penguin diagram such as the electroweak penguin and the penguin annihilation.}
$T$, the color-allowed external
$W$-emission tree diagram; $C$, the color-suppressed internal
$W$-emission diagram; $E$, the $W$-exchange diagram; $A$, the
$W$-annihilation diagram; $P$, the penguin diagram; and
$V$, the vertical $W$-loop diagram. It should be stressed
that these quark diagrams are classified according to the
topologies of weak interactions with all strong interaction
effects included and hence they are {\it not} Feynman graphs. All
quark graphs used in this approach are topological with all the
strong interactions included, i.e. gluon lines are included in all
possible ways. The diagrammatic approach was applied to hadronic $B$ decays
first in \cite{Chau91}. Various topological amplitudes have been extracted from
the data in \cite{Gronau94,Chiang,Chiang06,ChiangPV} after making some reasonable
approximations, e.g., SU(3) symmetry.

\section{EXPERIMENTAL TOOLS}
The experimental measurements involve separation of small samples
(10---2000 signal events) from total samples of \BB\ and
light quark-pairs (\qqbar) of several billion events.  The background is
typically dominated by the copious \qqbar\ production, where the
background after preliminary sample selection is often 1000 times
larger than the signal.  The separation of the signal from the large
background relies on a number of variables that are common to all
analyses.  For signal, the mass of a candidate, reconstructed from the
charged and neutral tracks in the event, is equal to the $B$ mass,
approximately 5.28 GeV/c$^2$.  The center-of-mass energy of the candidate
is equal to one-half of the $\Upsilon(4S)$ rest energy since all
$e^+e^-$ $B$ experiments take advantage of the increased production rate of
\BB\ events at this resonance.  Finally there are several
variables that take advantage of the difference in the shape of the
signal and background events.  In the center-of-mass system, signal is
typically spherical since the $B$ mesons are produced nearly at rest.
The \qqbar\ background events are are characterized by back-to-back ``jets"
of particles with small transverse momentum with respect to the
direction of the leading quarks.  For $B$ decays involving resonances in
the final state ($\rho$, $K^*$, $\eta$, \etapr, $\omega$, $\phi$, etc),
the mass of the daughter particles from the decay of these resonances is
also used.  This is useful because the background often is dominated by events
with combinatorial background, where the resonance candidate is not real but
rather is composed of combinations of particles from the \qqbar\ event
that happen to have an invariant mass near that of the resonance.  For
spin-1 (vector) particles, the so-called helicity angle is also often
useful since the decay of the daughters in the vector particle's rest
frame is not typically uniform.

Most analyses combine some of the above quantities into a maximum
likelihood (ML) fit.  Such a fit characterizes the signal and background with
probability density functions (PDFs) that describe the distribution
expected for each variable.  It is the difference between the shapes of
the signal and background PDFs that allow extraction of the signal from the
very large backgrounds.  The shape of these PDFs is typically determined
from Monte Carlo (MC) simulation for signal and \BB\ background events, with
checks provided by ``control" samples of other more copious decays.  The
data itself is used to determine the PDFs for the \qqbar\ background.
The free variables in the ML fit typically include the yields of the
signal and backgrounds, the value of \CP\ violation parameters where
relevant, the longitudinal polarization fraction for decays with two
particles with non-zero spin, and
often some of the parameters that determine the \qqbar\ background shapes.
The structure of the likelihood often assumes that the input observables
are uncorrelated.  This assumption is tested with data and MC and
correlations are typically below 10\%.  The residual correlations may
cause small signal biases ($\sim$10\%) which are evaluated with MC and
appropriate corrections are made.

\section{TWO-BODY DECAYS}

\subsection{Branching Fractions}
The general expressions of topological amplitudes are
 \begin{eqnarray} \label{eq:ampBpipi}
 A(\Bz\to\pi^+\pi^-) &=& T+P+{2\over 3}P^c_{\rm EW}+E+V, \nonumber \\
 A(\Bz\to\pi^0\pi^0) &=& -{1\over\sqrt{2}}(C-P+P_{\rm EW}+{1\over
3}P^c_{\rm EW}-E-V),\qquad\qquad  \\
 A(B^+\to \pi^+\pi^0) &=& {1\over\sqrt{2}}(T+C+P_{\rm EW}+P^c_{\rm
 EW}), \nonumber
 \end{eqnarray}
for tree-dominated $B\to\pi\pi$ decays,
\begin{eqnarray}
 A(\Bz\to K^+K^-)&=& E+P_A, \nonumber \\
 A(\Bz\to \Kz\Kzb)&=& P-{1\over 3}P_{\rm
EW}^c+P_A,\qquad\qquad\qquad\qquad\qquad\qquad\nonumber \\
 A(B^+\to K^+\Kzb) &=& A+P-{1\over 3}P_{\rm EW}^c,
\end{eqnarray}
for $B\to K\Kbar$ decays, and
 \begin{eqnarray} \label{eq:ampBKpi}
 A(\Bz\to K^+\pi^-) &=& P'+T'+{2\over 3}P'^c_{\rm EW}+P'_A, \nonumber \\
 A(\Bz\to \Kz\pi^0) &=& {-1\over \sqrt{2}}(P'-C'-P'_{\rm EW}-{1\over 3}P'^c_{\rm EW}+P'_A), \\
 A(B^+\to \Kz\pi^+) &=& P'-{1\over 3}P'^c_{\rm EW}+A'+P'_A,  \nonumber \\
 A(B^+\to K^+\pi^0) &=& {1\over\sqrt{2}}(P'+T'+C'+P'_{\rm
 EW}+{2\over 3}P'^c_{\rm EW}+A'+P'_A), \nonumber
 \end{eqnarray}
for $B\to K\pi$ decays,
where $P_{\rm EW}$ and $P^c_{\rm EW}$ are color-allowed and
color-suppressed electroweak penguin amplitudes, respectively, and
$P_A$ is the penguin-induced weak annihilation amplitude. We use unprimed and
primed symbols to denote $\Delta S=0$ and $|\Delta S|=1$ transitions.

\begin{figure}[htbp!]
\centerline{\psfig{figure=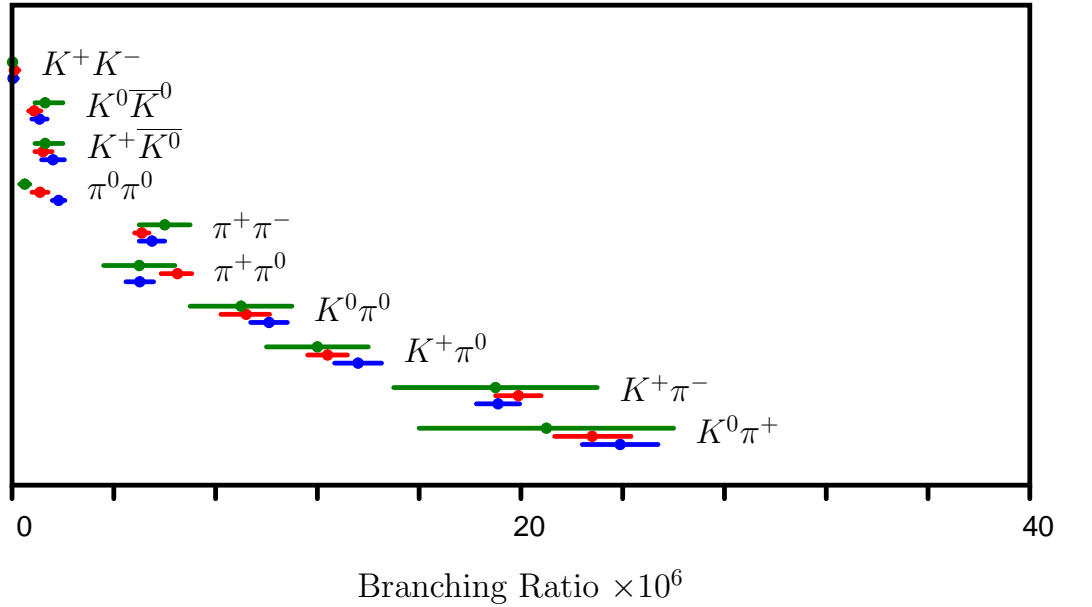,width=5.5in,clip=}}
\caption{Branching fraction measurements of 2-body decays from BABAR (blue)
\cite{2bodyBA1,2bodyBA2,2bodyBA3,2bodyBA4} and Belle (red)
\cite{2bodyBe1,2bodyBe2,2bodyBe3} and theoretical
predictions (green) \cite{Beneke03,LuShenWang,Li05,Zupan}.
}
\label{f:2-body}
\end{figure}

Experimental results for branching fractions from BABAR and Belle and
theoretical predictions are summarized
in Fig.~\ref{f:2-body}.  Here and in the following tables, the theoretical
values and errors are from weighted averages of the various predictions
with the errors divided by $\sqrt 3$ since the quoted theory errors indicate
a range as parameters are varied, not $1\sigma$ errors.

As mentioned in Sec. 2.2,  endpoint divergences will occur in the QCDF
approach in the penguin annihilation and hard spectator scattering amplitudes,
which are often parametrized as \cite{BBNS99}
\begin{eqnarray}
X_A=\ln\left({m_B\over \Lambda_h}\right)(1+\rho_A e^{i\phi_A}),
 \qquad
X_H=\ln\left({m_B\over \Lambda_h}\right)(1+\rho_H e^{i\phi_H}),
\end{eqnarray}
where the parameters $\rho_{A,H}$ and $\phi_{A,H}$ are real and $\Lambda_h$ is
a typical hadronic scale of order 500 MeV. Since these parameters
are unknown within QCD factorization, the central values of QCDF predictions
correspond to $\rho_{A,H}=0$ and the power corrections due to annihilation
effects give the largest theoretical uncertainties. Hence, for
penguin-dominated modes, the central values predicted by QCDF are usually
systematically below the measurements. Penguin annihilation and hard spectator
scattering amplitudes are calculable in the pQCD approach as the endpoint
singularity is overcome by the parton's transverse momentum. The predicted
central values by pQCD for penguin-dominated decays are normally higher than
those of QCDF.

The predicted $\pi^+\pi^-$ branching
fraction of $\sim7\times 10^{-6}$ is too large, whereas $\pi^0\pi^0$ of order
$0.3\times 10^{-6}$ is too small compared with experiment (the ratio of
the branching fractions is predicted much more precisely than one would
infer from Fig.~\ref{f:2-body}). If the penguin
amplitudes $P$, $P_{\rm EW}$, $P_A$ and the annihilation $E$ are neglected,
a fit to the ratio of $\Bz\to\pi^0\pi^0$ and $\Bz\to\pi^+\pi^-$ rates will yield
$|C/T|\sim 0.50-0.60$. In the short-distance ($SD$) factorization approach,
$|C/T|_{\rm SD}\sim 0.2-0.3$. It is thus a challenge to theorists to
understand $\Bz\to \pi^0\pi^0$ and $\pi^+\pi^-$ decays.

As for $B\to K\Kbar$ decays, $B^+\to K^+\Kzb$ and
$\Bz\to\Kz\Kzb$ are dominated by the CKM-suppressed $b\to d g$
penguin, while $\Bz\to K^+K^-$ proceeds only through weak annihilation. Hence,
the branching fraction is expected to be of order $10^{-6}$ for the former
and $10^{-8}$ for the latter, in agreement with the data.

The $B\to K\pi$ decays are dominated by penguin contributions because of
$|V_{us}V_{ub}^*|\ll |V_{cs}V_{cb}^*|\approx |V_{ts}V_{tb}^*|$ and the large
top quark mass. For the ratios defined by
\begin{eqnarray}
R_c\equiv {2\Gamma(B^+\to K^+\pi^0)\over \Gamma(B^+\to \Kz\pi^+)},
\qquad R_n\equiv {\Gamma(\Bz\to K^+\pi^-)\over 2\Gamma(\Bz\to \Kz\pi^0)},
\end{eqnarray}
we have $R_c=R_n$ if the other quark-diagram amplitudes are
negligible compared with $P'$. The current experimental measurements give
$R_c=1.12\pm0.07$ and $R_n=0.99\pm0.07$.
There are two approximate sum rules for $B\to K\pi$ rates \cite{Gronau99} and
rate asymmetries \cite{Atwood98}
\begin{eqnarray}
\Gamma(K^+\pi^-)+\Gamma(\Kz\pi^+)&\approx& 2[\Gamma(K^+\pi^0)+\Gamma(\Kz\pi^0)], \nonumber \\
\DG(K^+\pi^-)+\DG(\Kz\pi^+)&\approx& 2[\DG(K^+\pi^0)+\DG(\Kz\pi^0)],
\end{eqnarray}
based on isospin symmetry, where
$\DG(K\pi)\equiv \Gamma(\Bbar\to\bar K\bar\pi)-\Gamma(B\to K\pi)$.
The rate sum rule is fairly well satisfied by experiment.

\subsection{Direct \bma{\CP} Asymmetries}
It is well known that  it requires at least two amplitudes with nontrivial
relative strong and weak phases to produce direct \CP\ violation. The
direct \CP\ asymmetry observed in $\Bz\to K^+\pi^-$ decays,
\begin{eqnarray}
\acp(K^+\pi^-)\equiv{\Gamma(\Bzb\to K^-\pi^+)-\Gamma(\Bz\to K^+\pi^-)\over
\Gamma(\Bzb\to K^-\pi^+)+\Gamma(\Bz\to K^+\pi^-)}\label{eq:acpdef}
\end{eqnarray}
requires that the relative
strong phase $\delta$ between $t'=T'+P'^c_{\rm EW}$ and
$p'=P'-{1\over 3}P'^c_{\rm EW}+P'_A$ [see Eq.~(\ref{eq:ampBKpi})] be of order
$15^\circ$.  In QCDF, the phase $\delta$ induced perturbatively from vertex
corrections and penguin diagrams is too small and has a wrong sign \cite{Beneke03}.
Therefore, it is necessary to include the contribution from the long-distance
(LD) strong phase. The nonperturbative strong phase induced from the weak decay
$B\to D^{(*)}D_s^{(*)}$ followed by the final-state rescattering of
$D^{(*)}D_s^{(*)}\to K\pi$ can reproduce the experimental observation. In the
pQCD approach \cite{pQCD}, a large SD strong phase arises
from the penguin-annihilation diagram $P'_A$, though this is  in contrast to
the conventional wisdom that strong phases are basically nonperturbative in
nature.

\begin{figure}[htbp!]
\centerline{\psfig{figure=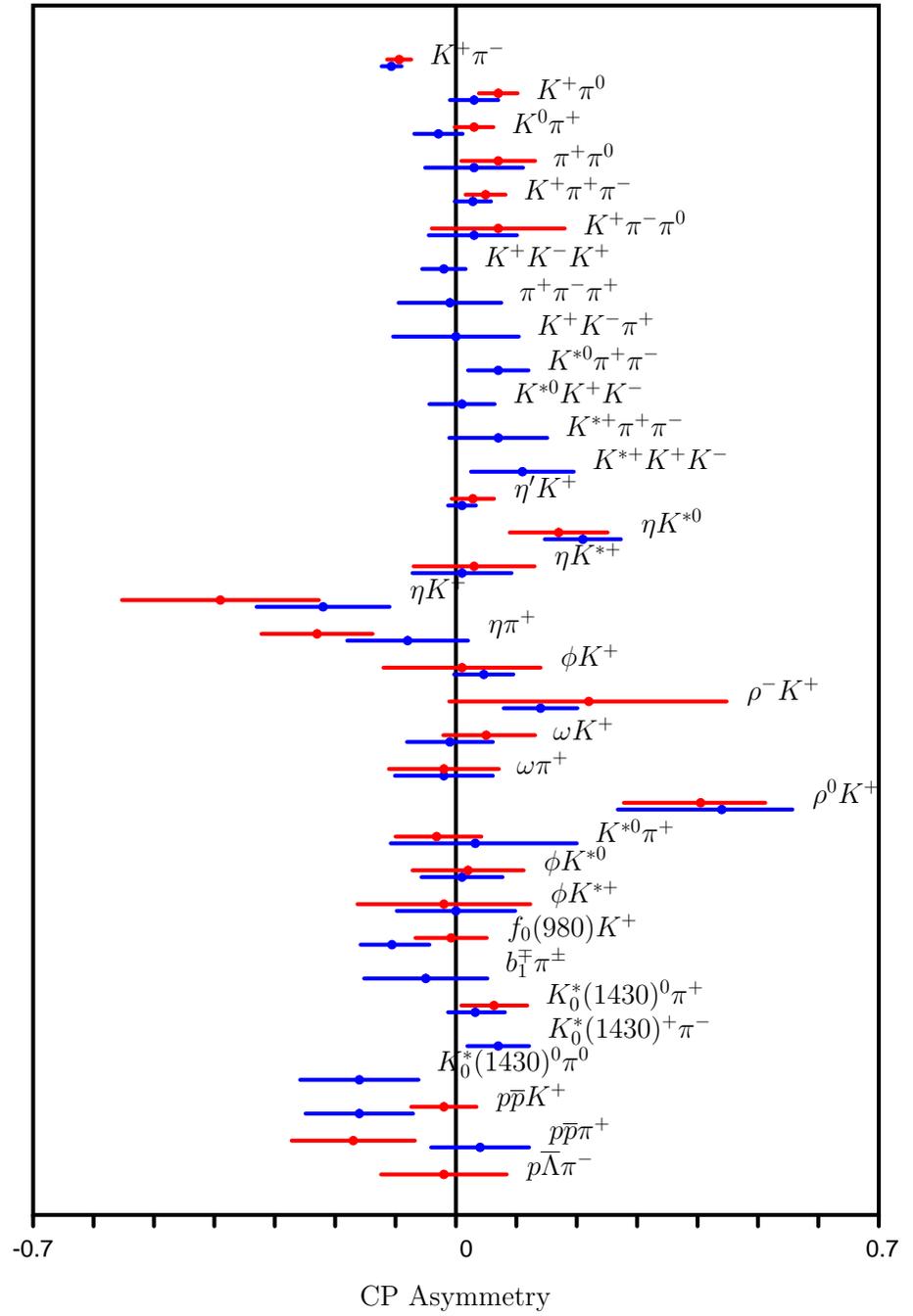,width=4.75in,clip=}}
\caption{A sample of the most precise direct $CP$ measurements from
BABAR (blue) and Belle (red). All of these measurements are obtained
from time-independent charge asymmetries.}
\label{f:acp}
\end{figure}

In Fig.~\ref{f:acp} we show \acp\ for the most precisely measured decays,
typically those with an average uncertainty $<$0.10 or a deviation from
zero of at least three standard deviations.  Of these only $\Bz\to
K^+\pi^-$ and $B^+\to\rho^0 K^+$ have world averages for \acp\ that are
more than 4 standard deviations from 0.  Theory predictions are not
included since the lack of knowledge of strong phases makes them very
uncertain.  All of these decays are measured from charge asymmetries as
suggested by Eq.~\ref{eq:acpdef}.  We defer until Sec.~\ref{sec:TD} a
discussion of the \CP\ asymmetries of \CP-eigenstate modes such as
$\Bz\to\pi^+\pi^-$ that are measured with time-dependent techniques.

If the color-suppressed $C'$, color-allowed electroweak penguin $P'_{\rm EW}$
and annihilation $A'$ are negligible compared with the tree amplitude $T'$, it is
clear from Eq. (\ref{eq:ampBKpi}) that the decay amplitudes of $K^+\pi^0$ and
$K^+\pi^-$ will be the same apart from a trivial factor of $1/\sqrt{2}$. Hence,
one will expect that $\acp(K^+\pi^0)\approx \acp(K^+\pi^-)$, while they
differ by 5.3 $\sigma$ experimentally,
$\Delta A_{K\pi}=\acp(K^+\pi^0)-\acp(K^+\pi^-)=+0.148\pm0.028$. Since
$A(B^+\to K^+\pi^0)\propto t'+c'+p'$ and $A(\Bz\to K^+\pi^-)\propto t'+p'$ with
$c'=C'+P'_{\rm EW}$, the puzzle is resolved provided that $c'/t'$ is of order
$0.5\sim 0.6$ with a negative relative phase. There are two possibilities for a
large $c'$: either a large color suppressed $C'$ or a large electroweak penguin
$P'_{\rm EW}$. We note that a global fit to $\pi\pi$, $K\pi$ and $K\bar K$
data in the diagrammatic approach gives a relative strong phase of
$(-56\pm10)^\circ$ between $C'$ and $T'$ \cite{Chiang06}. Various scenarios
for accommodating large $C'$ \cite{Li05} or $P'_{\rm EW}$ \cite{LargeEWP} have
been proposed. It is evident from the discussion above that a large
color-suppressed amplitude with a sizable relative phase to the tree amplitude
can solve both the $K\pi$ puzzle in \CP\ violation and the enhancement of
$\Bz\to\pi^0\pi^0$ in rate.

Based on SU(3) flavor symmetry, direct \CP\ asymmetries in $K\pi$ and
$\pi\pi$ systems are related as \cite{Deshpande}:
\begin{eqnarray}
\DG (K^+\pi^-)=-\DG (\pi^+\pi^-), \qquad \DG (\Kz\pi^0)=-\DG (\pi^0\pi^0).
\end{eqnarray}
The first relation leads to
$\acp(\pi^+\pi^-)=[{\cal B}(K^+\pi^-)/{\cal
B}(\pi^+\pi^-)]\acp(K^+\pi^-)\approx 0.37$\,, which is in good
agreement with the current world average of $0.38\pm0.06$.

\section{THREE-BODY DECAYS}

\begin{figure}[htbp!]
\centerline{\psfig{figure=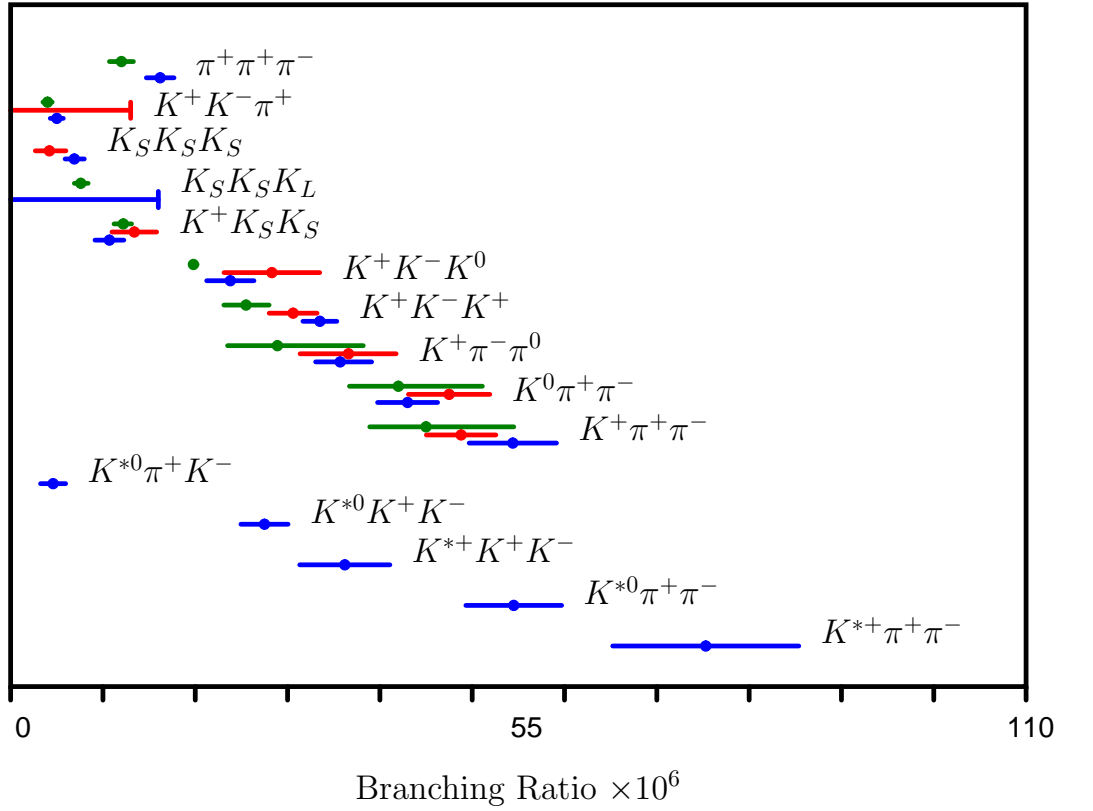,width=5.5in,clip=}}
\caption{Branching fraction measurements of 3-body decays from BABAR (blue)
\cite{3bodyBA1,3bodyBA2,3bodyBA3,3bodyBA4,3bodyBA5,3bodyBA6,3bodyBA7,3bodyBA8,3bodyBA9,3bodyBA10,3bodyBAA,3bodyBAB}
and Belle (red)
\cite{3bodyBe1,3bodyBe2,3bodyBe3,3bodyBe4} and theoretical
predictions (green) \cite{CCS3body}.  The predictions include both
resonant and nonresonant contributions.
}
\label{f:3-body}
\end{figure}

Three-body decays of heavy mesons are more complicated than the
two-body case as they receive both resonant and nonresonant
contributions. They are generally dominated by intermediate vector and scalar
resonances. The analysis of these decays using the Dalitz plot technique
enables one to study the properties of various resonances. Moreover, the
Dalitz plot analysis of 3-body $B$ decays provides a nice
methodology for extracting information on the unitarity triangle
in the standard model.
The experimental results are summarized in Fig.~\ref{f:3-body}.  The
agreement between Belle, BABAR, and theoretical predictions is good in
all cases.

\subsection{Resonant Contributions}
Under the factorization approximation, vector and scalar meson resonances
contribute to the two-body matrix elements $\langle P_1P_2|V_\mu|0\rangle$ and
$\langle P_1P_2|S|0\rangle$, respectively, with $S$ being a scalar density.
They can also contribute to the three-body matrix element
$\langle P_1P_2|V_\mu-A_\mu|B\rangle$.  Resonant effects are generally
described in terms of the usual Breit-Wigner formalism. For example,
 \begin{eqnarray}
 && \langle K^+(p_1)K^-(p_2)|\bar q\gamma_\mu q|0\rangle_R = \sum_i\langle K^+K^-|V_i\rangle\, {1\over m_{V_i}^2-s-im_{V_i}\Gamma_{V_i}}\langle V_i|\bar
q\gamma_\mu
q|0\rangle, \nonumber \\
&& \langle K^+(p_1)K^-(p_2)|\bar ss|0\rangle_R = \sum_i\langle K^+K^-|S_i\rangle\, {1\over m_{S_i}^2-s-im_{S_i}\Gamma_{S_i}}\langle S_i|\bar
ss|0\rangle,
 \end{eqnarray}
where $s=(p_1+p_2)^2$, $V_i=\phi,\rho,\omega,\cdots$ and
$S_i=f_0(980),f_0(1370),f_0(1500),\cdots$. Once the strong couplings for
$V_i(S_i)\to K^+K^-$ and the decay constants of $V_i$ and $S_i$ are known, we
are ready to compute various resonant contributions to the 3-body decays
of interest. Using the narrow width approximation for resonance $R$
 \begin{eqnarray} \label{eq:fact}
 \Gamma(B\to RP\to P_1P_2P)=\Gamma(B\to RP)\B(R\to P_1P_2),
 \end{eqnarray}
we can calculate the rates for the quasi-two-body decays $B\to PV$ and
$B\to SP$. For the details of theoretical calculations
of charmless 3-body decays of $B$ mesons based on the factorization approach,
see \cite{CCS3body}; for a theoretical overview, see \cite{ChengFPCP08}.

\subsection{Nonresonant Contributions}
One of the salient features of 3-body decays is the large nonresonant
contribution to penguin-dominated modes. It is known that the nonresonant
signal in charm decays is small, less than 10\% \cite{PDG}. On the contrary,
the nonresonant fraction is about $\sim$ 90\%
in $B\to KKK$ decays, $\sim (17- 40)\%$  in $B\to K\pi\pi$ decays (smaller in
the $K\pi\pi^0$ decay), and $\sim$ 14\% in the $B\to\pi\pi\pi$
decay. Hence, the nonresonant 3-body decays play an essential
role in penguin-dominated $B$ decays. Nonresonant amplitudes in charm decays
are usually assumed to be uniform in phase space. However, this is no longer
true in $B$ decays due to the large energy release in weak $B$ decays.
This makes the Dalitz plot analysis of nonresonant contributions more
difficult. While both BABAR \cite{3bodyBA5,3bodyBA9} and Belle \cite{3bodyBe2}
have adopted the parametrization
 \begin{eqnarray} \label{eq:ANR}
A_{\rm NR}=(c_{12}e^{i\phi_{12}}e^{-\alpha
s_{12}^2}+c_{13}e^{i\phi_{13}}e^{-\alpha
s_{13}^2}+ c_{23}e^{i\phi_{23}}e^{-\alpha s_{23}^2})(1+b_{\rm
NR}e^{i(\beta+\delta_{\rm NR})})
 \end{eqnarray}
to describe the non-resonant $B\to KKK$ amplitudes, they differ in the
analysis of the nonresonant component in $B\to K\pi\pi$ decays.

A detailed theoretical analysis in \cite{CCS3body} indicates two distinct
sources of nonresonant contributions: a small contribution from the tree
transition and a large one from the matrix
elements of scalar densities, e.g., $\langle K\Kbar|\bar ss|0\rangle$,
induced from the penguin transition. This explains the dominance
of the nonresonant background in $B\to KKK$ decays, the sizable nonresonant
fraction in $K^-\pi^+\pi^-$ and
$\Kzb\pi^+\pi^-$ modes and the smallness of nonresonant rates in
$B\to\pi\pi\pi$ decays.

Mixing-induced \CP\ asymmetries of 3-body decays will be discussed in Sec.~7 below.

\section{QUASI-2-BODY DECAYS}

\subsection{\bma{$B\to PP$}}

\begin{figure}[htbp!]
\centerline{\psfig{figure=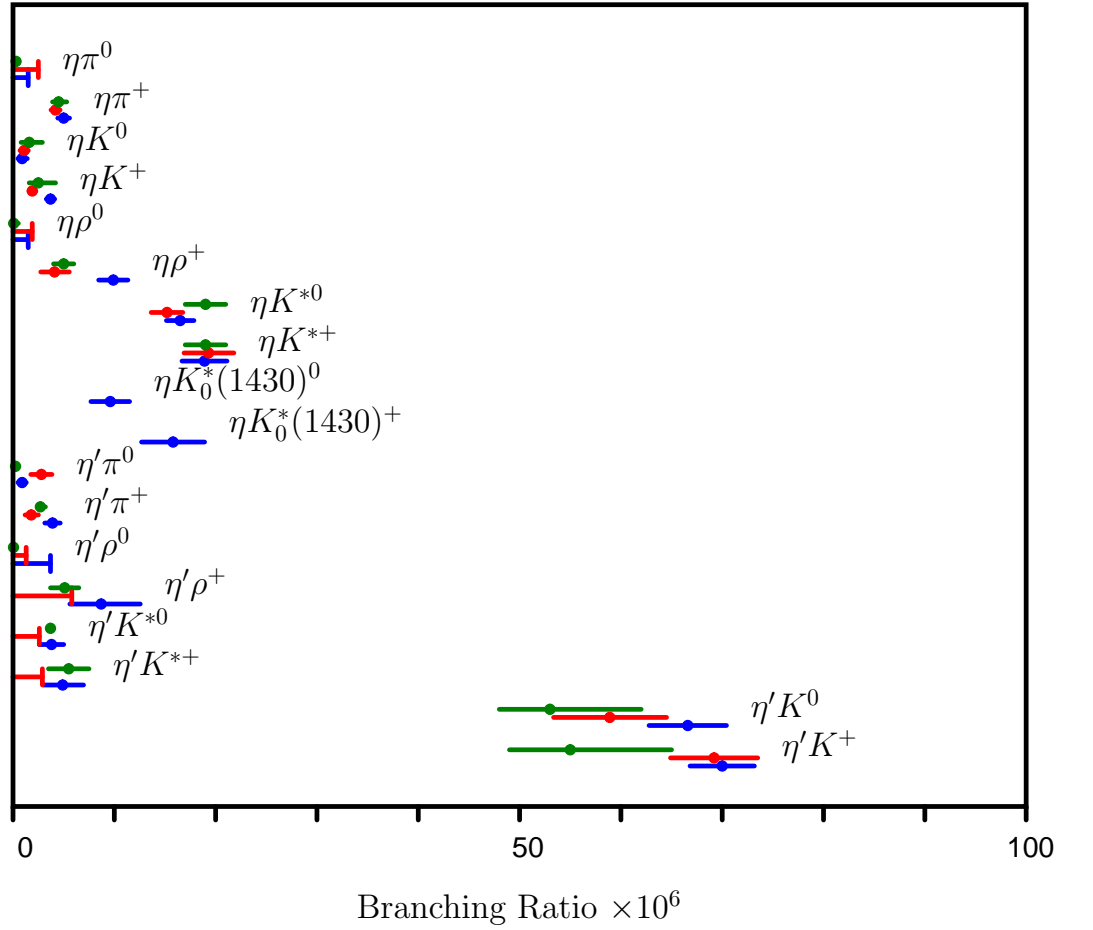,width=5.5in,clip=}}
\caption{Branching fraction measurements of decays with $\eta$ and $\etapr$
mesons from BABAR (blue) \cite{etaBA1,etaBA2,etaBA3,etaBA4,etaBA5,etaBA6}
and Belle (red) \cite{etaBe1,etaBe2,etaBe3,etaBe4,etaBe5} and theoretical
predictions (green) \cite{BenekeETA,Beneke03,Zupan,Wangpieta,Wang,Xiao}.
}
\label{f:eta}
\end{figure}

Among the 2-body $B$ decays, $B\to\etapr K$ has the largest branching fraction,
of order $70\times 10^{-6}$, while ${\cal B}(B\to\eta K)$ is only
$(1-3)\times 10^{-6}$. This can be roughly  understood as follows. Let's
express the \etapr\ and $\eta$ wave functions in the quark-flavor basis
$\eta_q=(u\bar u+d\bar d)/\sqrt{2}$ and $\eta_s=s\bar s$
\begin{eqnarray}
\eta=\cos\phi\eta_q-\sin\phi\eta_s, \qquad \etapr=\sin\phi\eta_q+\cos\phi\eta_s,
\end{eqnarray}
where the mixing angle is extracted from data to be $\phi=39.3^\circ$
\cite{FKS}. The interference between the $B\to\eta_q K$
amplitude induced by the $b\to sq\bar q$ penguin and the $B\to\eta_s K$
amplitude induced by $b\to ss\bar s$  is constructive for $B\to\etapr K$ and
destructive for $B\to\eta K$. This explains the large rate of the former and
the suppression of the latter. However, the predicted rates
are still somewhat smaller than the measurements (see Fig. \ref{f:eta}).

Many possible solutions to the puzzle for the abnormally large $\etapr K$ rate
have been proposed: (i) a significant flavor-singlet contribution
\cite{Chiang,Chiang06,BenekeETA}, (ii) a large $B\to \etapr$ form factor
\cite{Pham}, (iii) a contribution from the charm content of the $\etapr$,
(iv) an enhanced hadronic matrix element
$\langle 0|\bar s\gamma_5 s|\etapr\rangle$ due to the axial U(1) anomaly
\cite{Gerard}, (v) a large chiral scale $m_0^q$ associated with the $\eta_q$
\cite{Akeroyd}, (vi) a long-distance charming penguin in SCET \cite{Zupan}, and
(vii) a large contribution from the two-gluon fusion mechanism \cite{Ahmady}.

Because the \etapr\ is dominated by the flavor-singlet component $\eta_1$, it
is plausible that a sizable flavor-singlet amplitude $S'$ will account for the
large $\etapr K$ rates. When this contribution is included in the diagrammatic
approach, a global fit to the $B\to PP$ data leads to a flavor-singlet
amplitude $S$ about 3 to 4 times the magnitude of $P'_{\rm EW}$
\cite{Chiang,Chiang06}. This flavor-singlet contribution is also supported by
the consideration of the $B\to\eta^{(')}\pi^0$ decays. Their topological
quark decay amplitudes are
\begin{eqnarray}
A(B^+\to\eta\pi^+) = -{1\over \sqrt{3}}(t+c+2p+s), &&
A(\Bz\to\eta\pi^0) = -{1\over \sqrt{6}}(2p-s),  \nonumber \\
A(B^+\to\etapr\pi^+) = {1\over \sqrt{3}}(t+c+2p+4s), &&
A(\Bz\to\etapr\pi^0) = {1\over \sqrt{3}}(p+2s),
\end{eqnarray}
where $t=T+P_{\rm EW}^c$, $c=C+P_{\rm EW}$, $p=P-{1\over 3}P_{\rm EW}^c+P_A$,
$s=S-{1\over 3}P_{\rm EW}$, and we have assumed $\phi=35.3^\circ$ since
this value is algebraically simple.  Note that there is no tree contribution to
the $\eta^{(')}\pi^0$ modes. If $s=0$, one will naively expect that
$\B(\Bz\to\etapr\pi^0)={1\over 2}\B(\Bz\to\eta\pi^0)$.
QCDF predicts $\B(\Bz\to\eta\pi^0)=(0.28^{+0.48}_{-0.28})\times 10^{-6}$ and
$\B(\Bz\to\etapr\pi^0)=(0.17^{+0.33}_{-0.17})\times 10^{-6}$ \cite{Beneke03}.
The pQCD approach \cite{Wangpieta} has very similar results.
From the experimental measurements $\B(\Bz\to\eta\pi^0)<1.5\times 10^{-6}$ and
$\B(\Bz\to\etapr\pi^0)=(0.9\pm0.4)\times 10^{-6}$ by BABAR \cite{etaBA4} and
$\B(\Bz\to\etapr\pi^0)=(2.8\pm1.0)\times 10^{-6}$ by Belle \cite{etaBe2}, it
appears that the
current predictions of QCDF and pQCD may be too small for $\Bz\to\etapr\pi^0$.
In the presence of $S$, $\Bz\to\etapr\pi^0$ is enhanced:
$\B(\Bz\to\etapr\pi^0)\approx \B(\Bz\to\eta\pi^0)\sim 1.0\times 10^{-6}$
\cite{Chiang06}.

Since the two penguin processes $b\to ss\bar s$ and $b\to sq\bar q$ contribute
destructively to $B\to \eta K$, the penguin amplitude is comparable in
magnitude to the tree amplitude induced from $b\to us\bar u$, contrary to the
decay $B\to \etapr K$ which is dominated by large penguin amplitudes.
Consequently, a sizable direct CP asymmetry is expected in $B^+\to \eta K^+$
but not in $\etapr K^+$ \cite{BSS}. Indeed, the average of BABAR and Belle
measurements yields a 3$\sigma$ effect, $A_{\rm CP}(\eta K^+)=-0.27\pm0.09$\,.

\subsection{\bma{$B\to PV$}}
\begin{figure}[htbp!]
\centerline{\psfig{figure=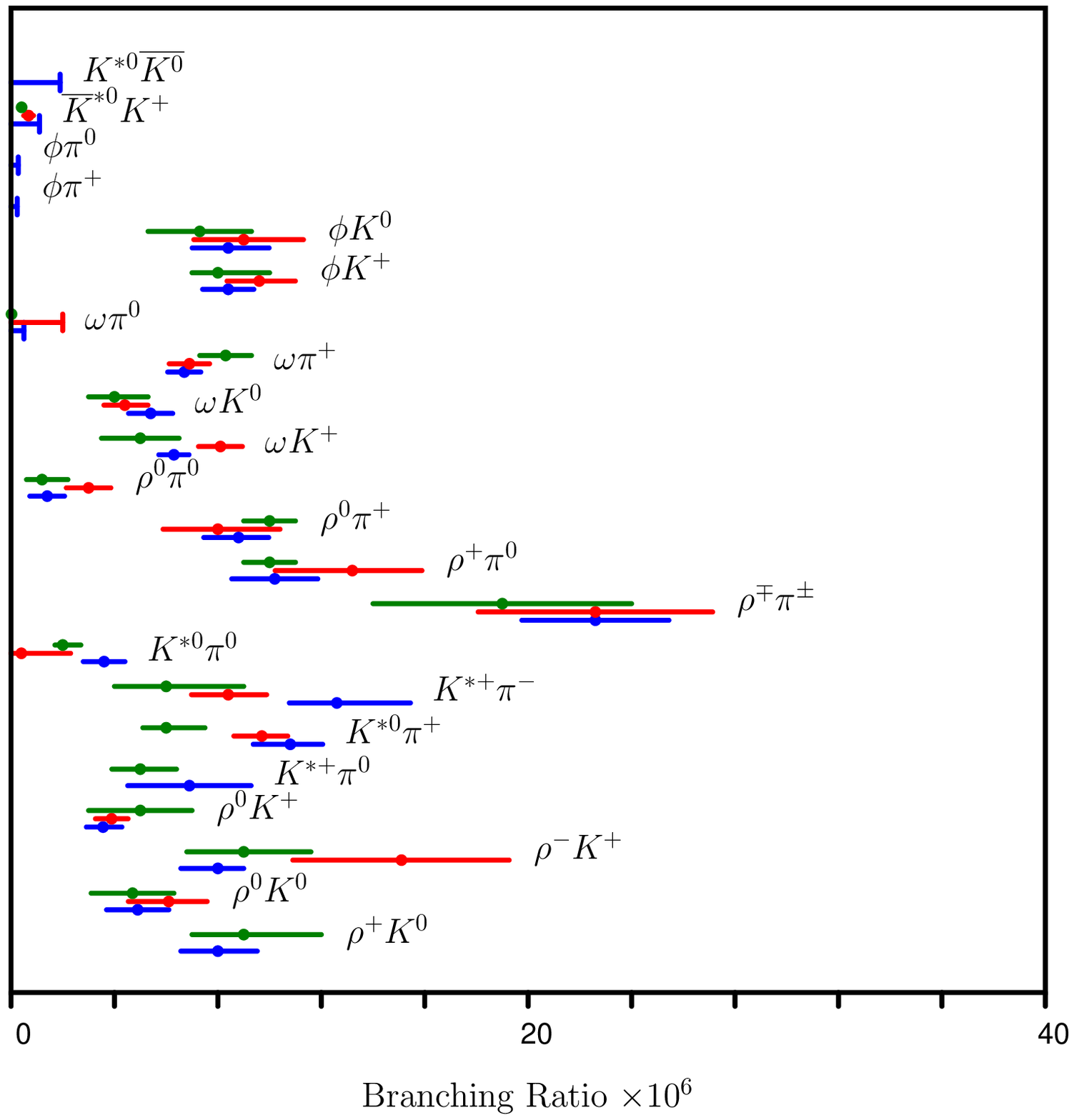,width=5.5in,clip=}}
\caption{Branching fraction measurements of $PV$ decays from BABAR (blue)
\cite{3bodyBA3,3bodyBA5,3bodyBAA,3bodyBAB,etaBA4,etaBA5,PVBA1,PVBA2,PVBA3,PVBA4,PVBA5,PVBA6,PVBA7,PVBA8,PVBA9,PVBAA}
and Belle (red)
\cite{3bodyBe2,3bodyBe3,3bodyBe4,PVBe1,PVBe2,PVBe3,PVBe4,PVBe5,PVBe6,PVBe7}
and theoretical predictions (green) \cite{Beneke03,Wang,Lu02,Liu,LGuo,DQGuo}.
}
\label{f:PV}
\end{figure}

The experimental results are summarized in Fig. \ref{f:PV}.
The decays $B\to PV$ have been studied in QCD-motivated approaches such as
QCDF, pQCD and SCET; see \cite{Wang} for comparison of the theory predictions
in various approaches  with experiment. For tree dominated decays,
the predictions of QCDF and pQCD are in agreement with experiment for
$\Bz\to\rho^\pm\pi^\mp$ and $B\to\omega\pi$, while the SCET calculations
are smaller due to the smallness of both $B\to\pi$ and $B\to\rho$ form factors
in SCET. On the contrary, SCET predicts larger rates than QCDF and pQCD for
the color-suppressed decays such as
$\Bz\to \rho^0\pi^0,~\eta^{(')}\rho^0,~\eta^{(')}\omega$ because the hard
form factor $\zeta_J^{PV}$ is comparable with the soft part
$\zeta^{PV}$ ($F^{PV}=\zeta_J^{PV}+\zeta^{PV}$) and is enhanced by a large
Wilson coefficient. The predicted $\B(\Bz\to \rho^0\pi^0)$, of order
$0.4\times 10^{-6}$ by QCDF \cite{Beneke03} and of order $0.1\times 10^{-6}$
by pQCD \cite{Lu02}, are too small compared with the experimental average of
$(2.0\pm0.5)\times 10^{-6}$ while the SCET calculations which rely on
some input from experiment are consistent with experiment \cite{Wang}.

For penguin-dominated $B\to PV$ decays, the predictions by QCDF are
systematically below the measurements \cite{Beneke03}, while the data can be
accommodated by fitting them to SCET.

As for $\acp(\Bz\to K^+\pi^-)$, QCDF predicts a wrong sign
for $\acp(B^+\to\rho^0K^+$).  The pQCD approach predicts too
large direct \CP\ violation in many $PV$ modes \cite{Wang}.

\subsection{\bma{$B\to VV$}}

\begin{figure}[htbp!]
\centerline{\psfig{figure=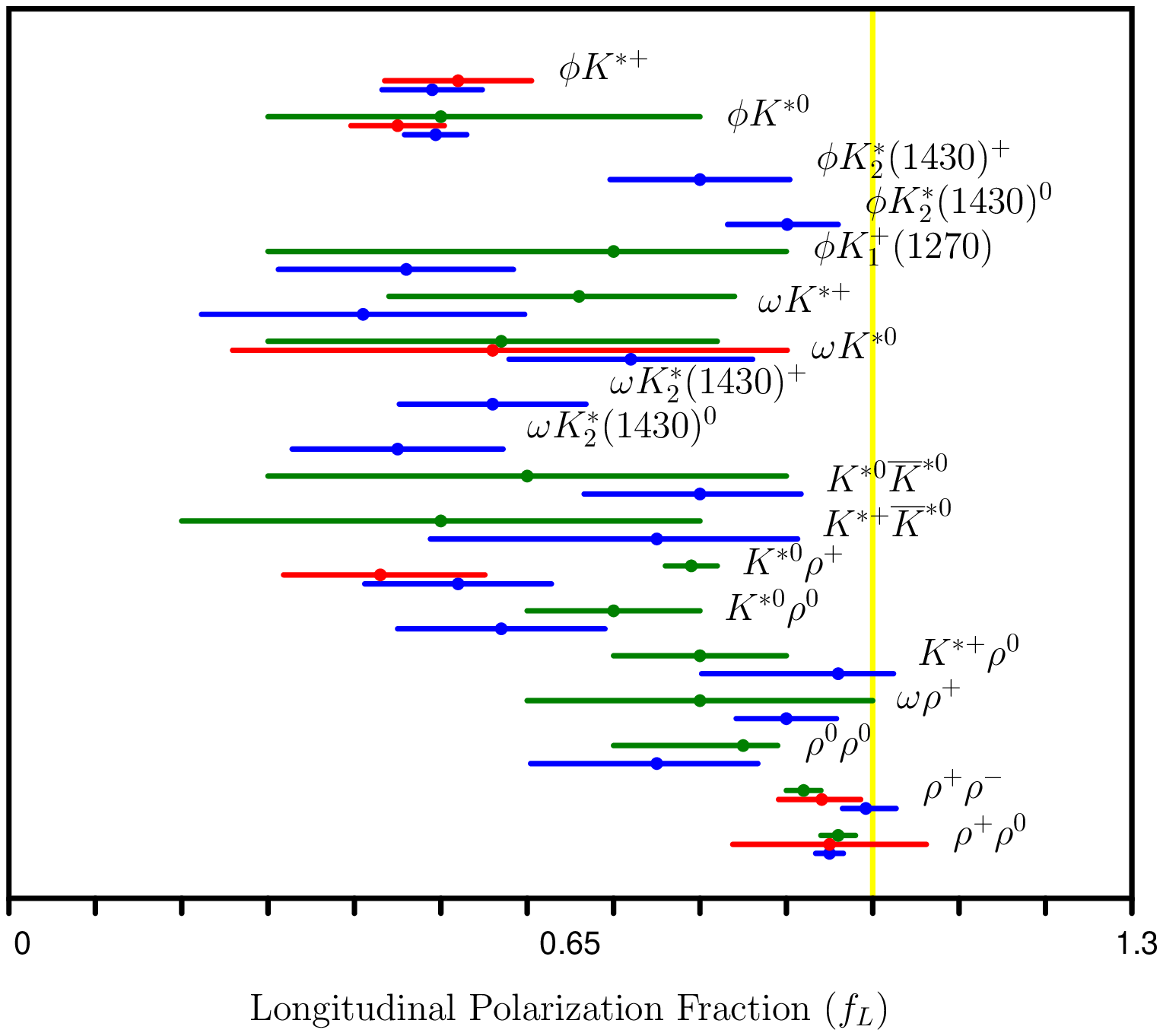,width=5.5in,clip=}}
\caption{Longitudinal polarization measurements from BABAR (blue)
\cite{polarBA1,VVBA1,VVBA2,VVBA3,VVBA4,VVBA5,VVBA6,VVBA8,polarBA2,VVBA9,VVBA10}
and Belle (red) \cite{PVBe2,VVBe1,VVBe2,VVBe3,VVBe4} and theoretical
predictions (green) \cite{Zhu,Huang,BenekeVV,Cheng08}.
}
\label{f:polar}
\end{figure}

The underlying mechanism for $B\to VV$ is more complicated than $PV$ and $PP$
modes as it involves three polarization vectors.
The decay amplitude of $B\to VV$ can be decomposed into three components, one
for each helicity of the final state: $A_0,A_+,A_-$. The transverse
amplitudes defined in  the transversity basis are related to the helicity ones
via
\begin{eqnarray}
 A_{\parallel}=\frac{A_++A_-}{\sqrt2}, \qquad
 A_{\bot}&=&\frac{A_+-A_-}{\sqrt2}.
 \label{eq:Atrans}
 \end{eqnarray}
It is naively expected that
the helicity amplitudes $A_h$ ($h=0,+,-$ ) for $B \to VV$ respect the
hierarchy pattern
$A_0:A_+: A_-=1:(\Lambda_{\rm QCD}/ m_b):(\Lambda_{\rm QCD}/m_b)^2$.
Hence, charmless $B\to VV$ decays are expected to be dominated by the
longitudinal polarization states and satisfy the scaling law,
 \begin{eqnarray} \label{eq:scaling}
1-\fL={\cal O}\left({m^2_V\over m^2_B}\right),
\qquad {f_\bot\over f_\parallel}=1+{\cal O}\left({m_V\over m_B}\right),
 \end{eqnarray}
with $\fL,~f_\bot$ and $f_\parallel$ being the longitudinal, perpendicular, and
parallel polarization fractions, respectively, defined by
\begin{eqnarray}
  f_\alpha\equiv \frac{\Gamma_\alpha}{\Gamma}
                     =\frac{|A_\alpha|^2}{|A_0|^2+|A_\parallel|^2+|A_\bot|^2},
\end{eqnarray}
with $\alpha=L,\parallel,\bot$. The experimental
measurements are summarized in Fig.~\ref{f:polar}.  The progress for
these measurements is quite impressive, with the uncertainty for a
handful of modes now $<0.05$\,.  In sharp contrast to the
$\rho\rho$ case, the large fraction of transverse polarization observed in
$B\to K^*\rho$ and $B\to\phi K^*$ decays at $B$ factories is a surprise
and poses an interesting challenge for
theoretical interpretations. Various mechanisms such as sizable
penguin-induced annihilation contributions \cite{Kagan},
\footnote{Historically, even before the experimental observation of the
polarization puzzle, it was already pointed out in \cite{CKL} that
penguin annihilation effects could reduce $\fL(\phi K^*)$ to 0.75.}
final-state
interactions \cite{Colangelo,CCSfsi}, form-factor tuning \cite{HNLiVV} and new
physics \cite{Newphysics} have been proposed for solving
the $B\to VV$ polarization puzzle.

Two recent calculations \cite{BenekeVV,Cheng08} indicate that NLO
nonfactorizable corrections from vertex and penguin corrections and hard spectator
scattering will render the positive-helicity amplitude of some $VV$ modes
comparable to the longitudinal one and hence will increase the transverse
polarization.  For example, \fL\ is naively expected to be
$1-4m_V/m_B^2\sim 0.90$ in $B\to\phi K^*$ and $\Bz\to K^{*0}\rho^0$
decays. However, NLO corrections decrease this expectation to
$\fL(\phi K^*)\sim 0.6$ and
$\fL(K^{*0}\rho^0)\sim 0.5$. Therefore, the polarization puzzle is alleviated
to a large extent by the consideration of NLO effects.  The theoretical
predictions in Fig.~\ref{f:polar} reflect these more recent calculations.

\begin{figure}[tp!]
\centerline{\psfig{figure=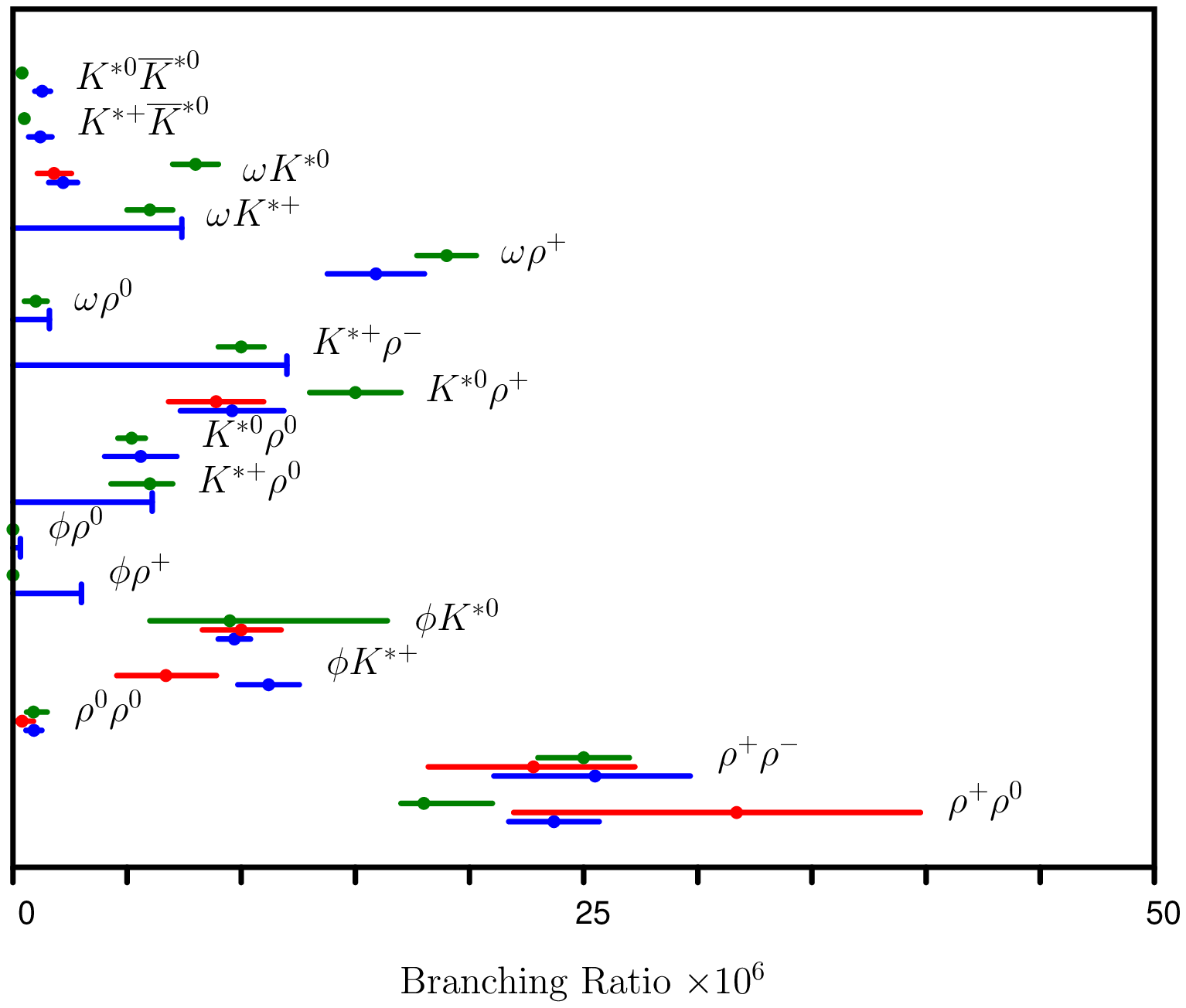,width=5.5in,clip=}}
\caption{Branching fraction measurements of $VV$ decays from BABAR (blue)
\cite{VVBA1,VVBA2,VVBA3,VVBA4,VVBA5,VVBA6,VVBA8,VVBA9,VVBA10} and Belle (red)
\cite{PVBe2,VVBe1,VVBe2,VVBe3,VVBe4,VVBe5} and theoretical
predictions (green) \cite{Zhu,Huang,LiMish,BenekeVV,Cheng08}.
}
\label{f:VV}
\end{figure}

According to the recent calculations based on QCDF \cite{Cheng08,BenekeVV},
there is a hierarchy pattern for the polarization fractions in $B\to K^*\rho$
decays:
\begin{eqnarray} \label{eq:KVrhofL}
\fL(K^{*+}\rho^0)> \fL(K^{*+}\rho^-)> \fL(K^{*0}\rho^+)> \fL( K^{*0}\rho^0).
\end{eqnarray}
This pattern is compatible with measurements though only two, $K^{*0}\rho^+$
and $K^{*0}\rho^0$, are well measured.  Improved measurements of all of these
decays are important in further testing the theoretical calculations.

Even though \fL\ can be substantially reduced  in
the presence of nonfactorizable corrections, the polarization anomaly is not
fully resolved unless the rate is also reproduced correctly. The experimental
branching fraction measurements are summarized in Fig.~\ref{f:VV}.  In
most cases the agreement between theory and experiment is quite good.
The recent measurements of the $B\to\omega K^*$ decays are well
below the predicted average though this is somewhat misleading since the
average is mostly from pQCD \cite{Huang}; the QCDF prediction
\cite{BenekeVV,Cheng08} is about a factor of
two smaller and in reasonable agreement with the data.  The
$B\to\phi K^*$ rate predicted by QCDF (pQCD) is too small (large) compared with
the data.  To improve the situation, QCD factorization and pQCD \cite{Mishima}
rely on penguin annihilation amplitudes, while SCET invokes charming penguins
\cite{SCET}, and the final-state interaction model considers final-state
rescattering of intermediate charm states \cite{Colangelo,CCSfsi}.

In QCDF, the theoretical model predicts
${\cal B}(\Bz\to\rho^0\rho^0)\sim 0.9\times 10^{-6}$ \cite{BenekeVV,Cheng08}
and ${\cal B}(\Bz\to\pi^0\pi^0)\sim 0.3\times 10^{-6}$ \cite{Beneke03},
whereas experimentally the latter has a rate larger than the former. One
plausible possibility is that final-state interactions are important for
$B\to \pi\pi$ but not for $B\to \rho\rho$. The $B\to \pi\pi$ amplitudes can
be decomposed into the $I=0$ and 2 isospin states with isospin phases
$\delta_0^\pi$ and $\delta_2^\pi$, respectively. When the isospin phase
difference is sizable, the $\pi^0\pi^0$ mode will be enhanced by the
final-state rescattering of $\pi^+\pi^-$ to $\pi^0\pi^0$ (this amounts to
enhancing the color-suppressed amplitude $C$, see also \cite{Chua}). Since $B\to \rho^+\rho^-$ has a
rate much larger than $B\to \pi^+\pi^-$, it is natural to expect that
$B\to \rho^0\rho^0$ will receive large enhancement via isospin final-state
interactions. The fact that the branching fraction of this mode is rather small
and is consistent with the theory prediction implies that the isospin phase
difference of $\delta_0^\rho$ and $\delta_2^\rho$ must be negligible and so is
the final-state interaction \cite{Vysotsky}.

\subsection{\bma{$B\to (S,~A,~T)M$}}

\begin{figure}[!]
\centerline{\psfig{figure=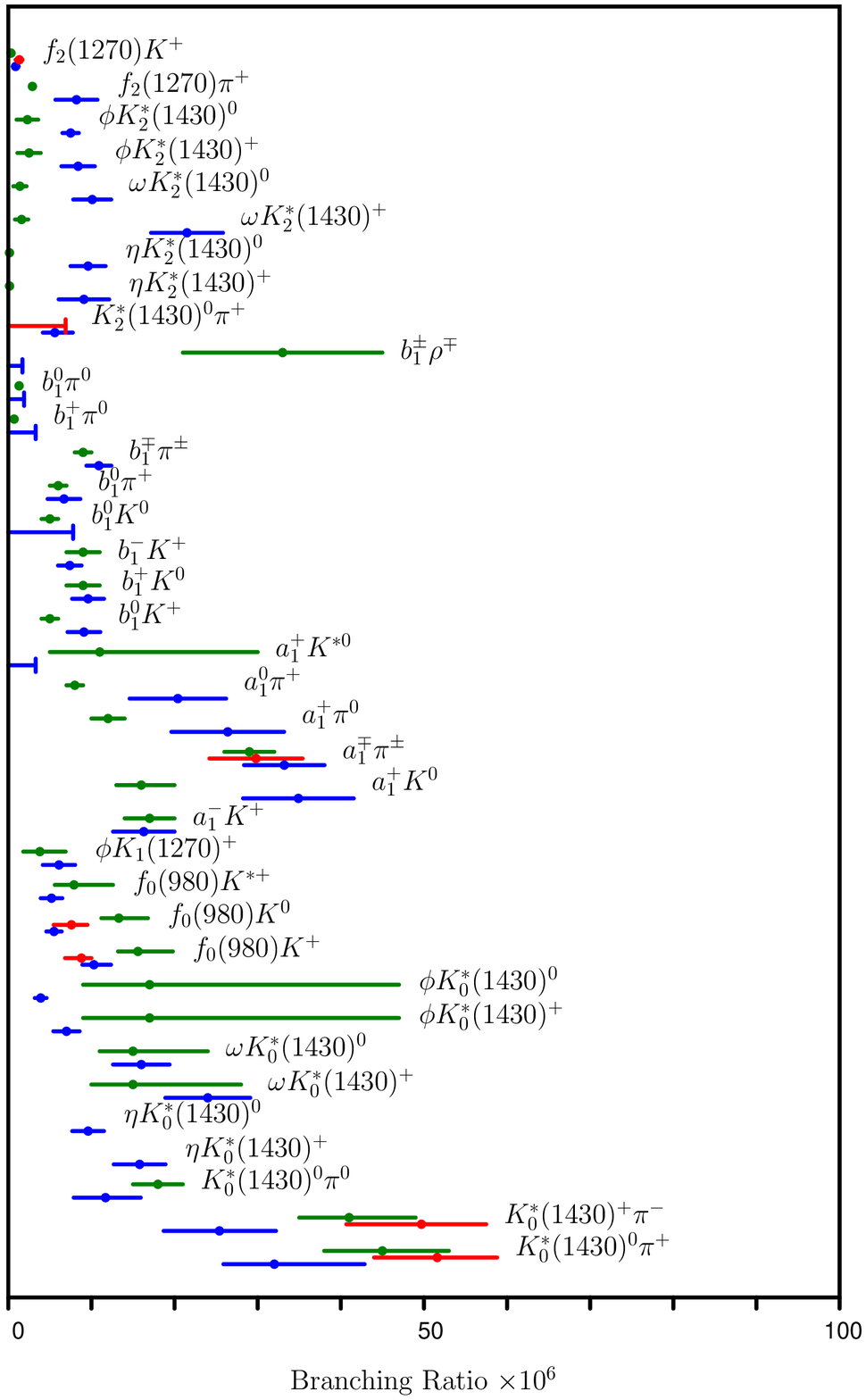,width=4.2in,clip=}}
\caption{Branching fraction measurements of decays involving $0^+$ (bottom
group), $1^+$ (middle group), and $2^+$ mesons (top group) from BABAR (blue)
\cite{3bodyBA4,3bodyBAA,3bodyBAB,etaBA1,VVBA2,VVBA9,polarBA2,otherBA1,otherBA2,otherBA3,otherBA4,otherBA5,otherBA6}
and Belle (red) \cite{3bodyBe2,3bodyBe3,3bodyBe4,otherBe1} and theoretical
predictions (green)
\cite{KimVV,ChengSP,WangShen,ChengAP,Shen,ChengSV,Cheng08,WangLiLu}.
}
\label{f:other}
\end{figure}

Much less is known about scalar, axial-vector and tensor mesons, though
there are already dozens of measurements involving these modes as decay
products of $B$ mesons.  In this section we summarize the experimental
measurements (see Fig.~\ref{f:other}) and the theoretical expectations.

\subsubsection{\bma{$B\to SP,SV$}} It is known that the identification of light
scalar mesons is difficult experimentally and the underlying structure of
scalar mesons is not well established theoretically.  It is hoped that through
the study of $B\to SP$, old puzzles related to the internal structure and
related parameters, e.g. the masses and widths, of light scalar mesons can
receive new understanding. For example, it has been shown \cite{ChengSP}
that if $a_0(980)$ is a $q\bar q$ bound state, the predictions are
$\B(\Bz\to a_0^\pm(980)\pi^\mp)\sim 8.2\times 10^{-6}$ and
$\B(\Bz\to a_0^-(980)K^+)\sim 4.3\times 10^{-6}$.  These exceed the current
experimental 90\% confidence level (C.L.) upper limits $3.1\times 10^{-6}$
and $1.9\times 10^{-6}$ \cite{etaBA2}, respectively, suggesting that the
four-quark nature for the $a_0(980)$ is favored.

One of the salient features of the scalar meson is that its vector decay
constant $f_S$ defined by
$\langle S(p)|\bar q_2\gamma_\mu q_1|0\rangle= f_S p_\mu$
is either zero or small (of order $m_d-m_u$, $m_s-m_{d,u}$).
Therefore, when one of the pseudoscalar mesons in
$B\to PP$ decays is replaced by the corresponding scalar, the resulting decay
pattern could be very different. For example, it is expected that
$\Gamma(B^+\to a_0^+\pi^0)\ll \Gamma(B^+\to a_0^0\pi^+)$ and
$\Gamma(\Bz\to a_0^+\pi^-)\ll \Gamma(\Bz\to a_0^-\pi^+)$ as the
factorizable contribution proportional to the decay constant of the scalar
meson is suppressed relative to the one proportional to the pseudoscalar
meson decay constant. This feature can be checked experimentally.

The decay $B\to f_0(980)K$ is the first charmless $B$ decay into a scalar meson
observed at $B$ factories \cite{Bellef0K}. It receives two different types of
penguin contributions: one from $b\to su\bar u$ and the other from
$b\to ss\bar s$. Due to the large scalar decay constant $\bar f^s_{f_0}$ of
order 370 MeV defined by $\langle f_0|\bar s s|0\rangle=m_{f_0}\bar f^s_{f_0}$ that appears in the penguin amplitude,
this decay is dominated by the $b\to ss\bar s$ penguin contribution with
predictions of about $15\times10^{-6}$ \cite{ChengSP}.  The experimental
measurements are for $\B(B\to f_0(980)K)\times\B(f_0(980)\to\pi^+\pi^-)$.
This product is of order $(5-10)\times 10^{-6}$ for the $B^+$ and $\Bz$ decays.
The theoretical predictions are consistent with experiment provided that
$\B(f_0(980)\to\pi^+\pi^-)\sim 0.50$.

\subsubsection{\bma{$B\to AP$}}
There are two distinct types of parity-even axial-vector mesons,
namely, $^3P_1$ and $^1P_1$. The $^3P_1$ nonet consists
of $a_1(1260)$, $f_1(1285)$, $f_1(1420)$ and $K_{1A}$, while the
$^1P_1$ nonet has $b_1(1235)$, $h_1(1170)$, $h_1(1380)$ and
$K_{1B}$. The physical mass eigenstates $K_1(1270)$ and
$K_1(1400)$ are mixtures of $K_{1A}$ and $K_{1B}$ states owing to
the mass difference of the strange and non-strange light quarks.

A prominent feature of the $^1P_1$ axial vector meson
is that its axial-vector decay constant is small, vanishing in the
SU(3) limit. This feature was confirmed by the BABAR observation
\cite{otherBA2} that $\Gamma(\Bz\to b_1^+\pi^-)\ll \Gamma(\Bz\to b_1^-\pi^+)$.
By contrast, it is expected that
$\Gamma(\Bz\to a_1^+\pi^-)\gg\Gamma(\Bz\to a_1^-\pi^+)$ since
$f_{a_1}\gg f_\pi$.

The predicted branching fractions for the $b_1K$ and $b_1\pi$ modes are
in good agreement with the BABAR measurements. The comparison for
$a_1^\pm\pi^\mp$ and $a_1^-K^+$ is sometimes good and sometimes not so good
(see Fig.~\ref{f:other}); improved measurements are needed.

\subsubsection{\bma{$B\to VA,AA$}}
Decays to $VA$ and $AA$ final states have been systematically studied within
the framework of QCD factorization \cite{Cheng08}. The calculations
indicate that some of the tree-dominated $a_1\rho$ and $b_1\rho$ modes have
sizable rates. For example,
$\B(\Bz\to a_1^\pm \rho^\mp)\sim 60\times 10^{-6}$ and
$\B(\Bz\to b_1^\pm\rho^\mp)\approx \B(\Bz\to b_1^-\rho^+)\sim 32\times 10^{-6}$.
Likewise, the $AA$ modes such as  $a_1^+a_1^-$, $a_1^+a_1^0$,
$a_1^+b_1^-$ and $a_1^+b_1^0$ are expected to have branching ratios of
$(20- 40)\times 10^{-6}$.  Of these, there are only a few upper limits
from BABAR.  The preliminary result for $\B(\Bz\to b_1^\pm\rho^\mp)$ is
that the branching fraction is $<1.7\times 10^{-6}$ (90\% C.L.).  The strong
disagreement with the theoretical prediction is not understood.

A comparison of theory with the current data on $B^+\to\phi K_1(1270)^+$,
$B^+\to\phi K_1(1400)^+$ \cite{polarBA2} and $B^+\to a_1^+K^{*0}$
\cite{otherBA6} seems to imply that penguin annihilation is small
in penguin-dominated $B\to VA$
decays.  The prediction of $\fL(\phi K_1(1270)^+)\approx 0.44$ in the absence of
penguin annihilation \cite{Cheng08} agrees well with the experimental
result of $\fL(\phi K_1(1270)^+)=0.46^{+0.12}_{-0.15}$ \cite{polarBA2}. This
indicates that it is the NLO correction rather than penguin annihilation
that is responsible for pushing the longitudinal polarization fraction in
$B^+\to\phi  K_1(1270)^+$ down to the level of 0.5\,.

\subsubsection{\bma{$B\to TP,TV$}}
Many charmless $B$ decays with a tensor meson in the final state have been
observed at $B$ factories (see Fig. \ref{f:other}). Moreover, BABAR has
measured \fL\ in the decays $B\to\phi K_2^*(1430)$ and $B\to\omega K_2^*(1430)$.
Contrary to the penguin-dominated $VV$ modes such as $\phi K^*$ and $\rho K^*$,
$\phi K_2^*(1430)$ has $\fL\sim0.85$.  Intriguingly, the
$B\to\omega K_2^*(1430)$ modes have \fL\ consistent with 0.5.  So far there
are only two theoretical studies of the charmless $B$ decay to a tensor meson,
both done with the generalized factorization approach \cite{KimVV}.
Neither of these calculations has a prediction for \fL.  At the moment,
the data for decays to tensor mesons is well ahead of the theory.

\section{BARYONIC \bfB\ DECAYS}

\subsection{Experimental Status}
\begin{figure}[htbp!]
\centerline{\psfig{figure=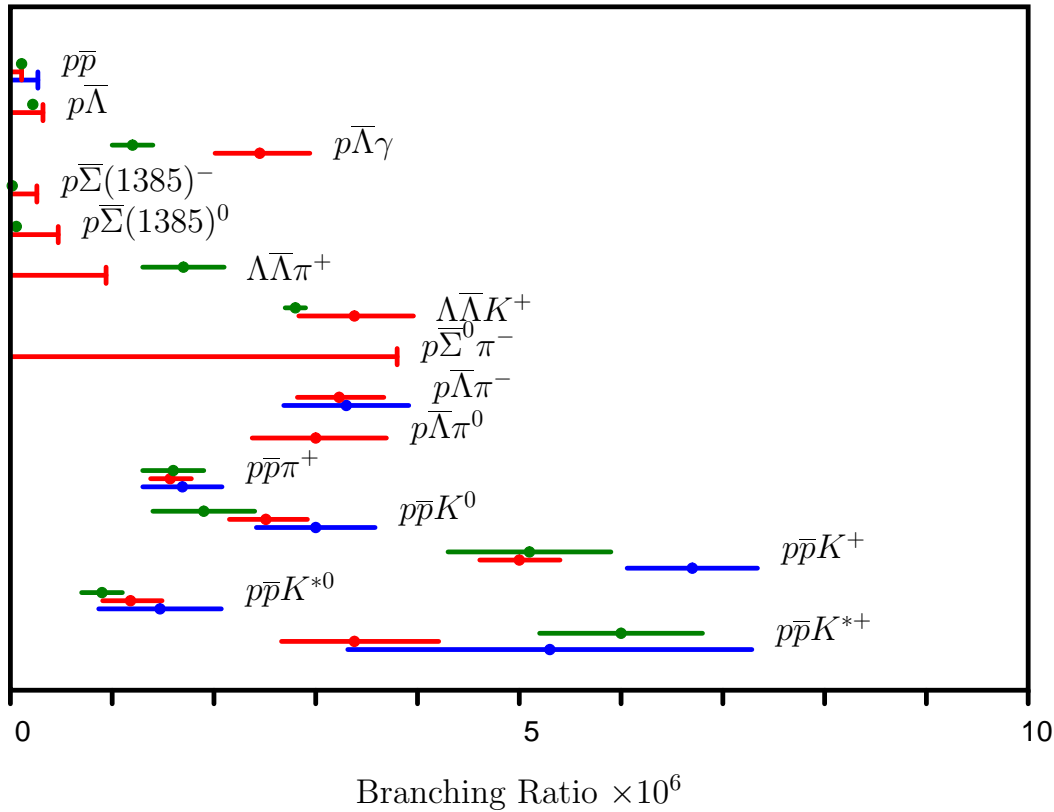,width=5.5in,clip=}}
\caption{Experimental results for decays with baryons from BABAR (blue)
\cite{baryBA1,baryBA2,baryBA3,baryBA4} and Belle (red)
\cite{baryBe1,baryBe2,baryBe3,baryBe4,baryBe5,baryBe6} and theoretical
predictions (green) \cite{ChengBary02,Hsiao}.
}
\label{f:bary}
\end{figure}

The experimental results for 15 decays involving light baryons are summarized
in Fig. \ref{f:bary}.  In most cases, the agreement between experiment
and theory is good.

\subsection{Threshold Enhancement}
A peak near the threshold area of the dibaryon
invariant mass spectrum has been observed in many 3-body baryonic $B$ decays.
The so-called threshold effect indicates that the $B$ meson is preferred to
decay into a baryon-antibaryon pair with low invariant mass accompanied by a
fast recoil meson. Threshold enhancement was first conjectured by Hou and Soni
\cite{HouSoni}. They argued that in order to have substantial branching fractions
for baryonic $B$ decays, one has to reduce the energy
release and  at the same time allow for baryonic ingredients to be present in
the final state. This is indeed the near-threshold effect mentioned above.

While various theoretical ideas \cite{HouSoni,Chengreview}
have been put forward to explain the low mass threshold
enhancement, this effect can be understood in terms of a simple
short-distance picture \cite{Suzuki}. In the two-body decays, one energetic
$q\bar q$ pair must be emitted back to back by a hard gluon in order to
produce a baryon and an antibaryon. Since this hard gluon is
highly off mass shell, the two-body decay amplitude is
suppressed by order of $\alpha_s/q^2$. In the three-body baryonic $B$ decays,
a possible configuration is that the baryons in the pair $\B_1\bar \B_2$ are
collinear and in the opposite direction from the meson. At the quark level,
the quark and antiquark emitted from a gluon are moving in nearly the same
direction. Since this gluon is close to its mass shell, the
corresponding configuration is not subject to the short-distance
suppression. This implies that the dibaryon pair tends to
have a small invariant mass.

\subsection{Two-body and Three-body Decays}
None of the two-body
charmless baryonic $B$ decays have been observed so far and the
present limit on their branching ratios has been pushed to the level of
$10^{-7}$ for $B\to p\bar p$ and $\Lambda\bar p$ \cite{baryBA1,baryBe2}.
The fact that three-body final states have rates larger than their
two-body counterparts, i.e., $\Gamma(B\to \B_1 \bar \B_2
M)>\Gamma(B\to \B_1\bar \B_2)$ is due to the threshold effect discussed above.


The study of 3-body decays is more complicated. The factorizable contributions
fall into two categories: (i) the transition process with a meson emission,
$\langle M|(\bar q_3 q_2)|0\rangle\langle \B_1\Bb_2|(\bar q_1b)|\Bbar\rangle$,
and (ii) the current-induced process governed by the factorizable amplitude
$\langle\B_1\Bb_2|(\bar q_1 q_2)|0\rangle\langle M|(\bar q_3 b)|\Bbar\rangle$.
The interested reader is referred to \cite{Chengreview} for further details.

\subsection{Radiative Decay}
Naively it appears that the bremsstrahlung process will lead to
$\Gamma(B\to\B_1\Bb_2\gamma)\sim {\cal O}(\alpha_{\rm
em})\Gamma(B\to\B_1\Bb_2)$, with $\alpha_{\rm em}$ being an
electromagnetic fine-structure constant, and hence the radiative
baryonic $B$ decay is further suppressed than the two-body
counterpart, making its observation very difficult at the present
level of sensitivity for $B$ factories. However, there is an
important short-distance electromagnetic penguin transition $b\to s \gamma$.
Because of the large top quark mass, the amplitude of
$b\to s\gamma$ is neither quark-mixing nor loop suppressed.
Moreover, it is largely enhanced by QCD corrections. As a
consequence, the short-distance contribution due to the
electromagnetic penguin diagram dominates over the bremsstrahlung.
The relatively large predictions of order $1\times 10^{-6}$
\cite{Chengrad02,Chengrad06,Gengrad} have been confirmed with a measurement
from Belle for the decay $B^+\to p\bar \Lambda\gamma$ \cite{baryBe2}.


\subsection{Angular Distribution}
Measurement of angular distributions in the dibaryon rest frame will
provide further insight to the underlying dynamics. The SD
picture and the pole model both predict a stronger correlation of the outgoing
meson with the baryon than the antibaryon in the decay $B\to
\B_1\Bb_2M$. This feature has been confirmed for $B^+\to p\bar
p\pi^+$ and $B^+\to p\bar\Lambda\gamma$, but not for $B^+\to p\bar pK^+$.
Both BABAR \cite{baryBA2} and Belle \cite{Belle:ppK} found that the $K^+$
in the latter decay prefers to be collinear with the $\bar p$
in the $p\bar p$ rest frame, contrary to the above expectation. This angular
correlation puzzle indicates that either some long-distance effects enter and
reverse the angular dependence or the dibaryon pair $p\bar p$ is produced from
some intermediate state e.g. baryonium.

Recently Belle has made a new measurement of the angular distribution of
$B^+\to p\bar\Lambda\pi^+$ \cite{baryBe3}. Naively, it is expected that the
pion has no preference for its correlation with the $\bar \Lambda$ or the
proton as the dibaryon
picks up energetic $s$ and $\bar u$ quarks, respectively, from the
$b$ decay. However, the new Belle measurement indicates a correlation between
the pion and the $\bar\Lambda$. In short, the correlation enigma has been found
in the penguin-dominated modes $B^+\to p\bar pK^+$ and
$B^+\to p\bar\Lambda\pi^+$ and it cannot be explained by the SD $b\to sg$
picture. This poses a great challenge to theorists.

\section{TIME-DEPENDENT \bma{\CP} VIOLATION}\label{sec:TD}

The $B$ Factories at KEK in Japan and PEP-II in California have
asymmetric energies: the electron beam has an energy of $8-9$ GeV
while the positron beam is about 3 GeV.  This asymmetry means that the
center of mass of the $\Upsilon(4S)$ is moving so that the produced \Bz\
and \Bzb\ mesons do not decay at the same point.  Belle and BABAR take
advantage of this to measure, with a precision of about 100$\,\mu$m, the
distance between the ``signal" $B$ decay and the ``tagged" $B$ decay.
The signal final state can come from either \Bz\ or \Bzb; the tagged $B$
is either a \Bz\ or \Bzb\ with the flavor determined primarily by the charge
of leptons or kaons in the event.   This measurement, together
with the known properties of the motion of the $\Upsilon(4S)$ system, allow
measurement of the (signed) time difference \dt\ between the \Bz\
and \Bzb\ decays.  Then the time-dependent asymmetry of the decays to a
final state $f$ is measured:
 \begin{eqnarray}
 {\Gamma(\Bbar(\dt)\to f)-\Gamma(B(\dt)\to f)\over
 \Gamma(\Bbar(\dt)\to f)+\Gamma(B(\dt)\to
 f)}=S_f\sin(\Delta m\dt)-C_f\cos(\Delta m\dt),
 \end{eqnarray}
where $\Delta m$ is the mass difference of the two neutral $B$
eigenstates, $S_f$ monitors mixing-induced \CP\ asymmetry and
$C_f$ measures direct \CP\ violation (Belle uses $A_f=-C_f$).

\begin{figure}[htbp!]
\centerline{\psfig{figure=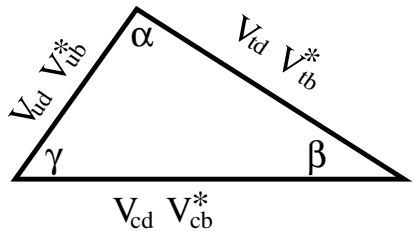,height=3cm}}
\caption{The CKM triangle, showing the angles $\alpha~(\phi_2),~\beta~(\phi_1)$,
and $\gamma~(\phi_3)$.}
\label{fig:UT}
\end{figure}

In 2001, BABAR and Belle used this technique to observe \CP\ violation
in the $B$ meson system for the first time, measuring the value of $S$
for the decay $B\to J/\psi K^0$ and similar $b\to c\bar c s$ decays.
Since there is only one quark-level process involved in these decays,
this is known to measure $\sin2\beta$ with an uncertainty of $\sim$0.001,
where $\beta$ (also called $\phi_1$) is one of the angles of the
CKM triangle (See Fig.~\ref{fig:UT}).  The ambiguity in extracting
$\beta$ is resolved with other measurements so that the current world
average is $\beta=(21\pm1)^\circ$.
In this section we discuss how charmless hadronic $B$ decays
can be used to measure the CKM angle $\alpha$ and to search for physics
beyond the Standard Model.

\subsection{Measurements of \bma{$\alpha$}}
Measurements of $S$ in the decay $\Bz\to\pi^+\pi^-$ can be used to extract the
angle $\alpha$.  However there are complications since the penguin amplitudes
are not small (``penguin pollution"), so a second weak phase is introduced.
Gronau and London showed that an isospin analysis \cite{GronLon} involving
also the decays $\Bz\to\pi^0\pi^0$ and $B^+\to\pi^+\pi^0$ can be used to
determine the amount of penguin pollution and hence extract the angle $\alpha$.

While the measurements of $S$ and $C$ for the decay $\Bz\to\pi^+\pi^-$
are fairly precise ($\delta S=0.07$) \cite{2bodyBA4,Bellepipi}, the
sensitivity to $\alpha$ is rather
poor since the penguin pollution is large in this decay.  The problem can be
seen from the relatively large branching fractions for the penguin-dominated
$B\to K\pi$ decays compared with $B\to\pi\pi$ (Fig.~\ref{f:2-body}).
The situation is reversed for the $VV$ decays where $B\to K^*\rho$ tends
to be smaller than $B\to\rho\rho$ (Fig.~\ref{f:VV}).  Thus the most
sensitive measurements of $\alpha$ have come from measurements of $S$
for the decay $\Bz\to\rho^+\rho^-$ \cite{VVBA4,Bellerhorho} and the
isospin analysis involving also
$\Bz\to\rho^0\rho^0$ and $B^+\to\rho^+\rho^0$.  Some complications arise
for these decays due to polarization and potential corrections to the isospin
analysis due to $\rho-\omega$ mixing, electroweak-penguin amplitudes,
and other isospin-breaking amplitudes.  The magnitude of all of these
effects is small compared with the present precision.  While the exact
numbers differ due to the statistical analysis used in the treatment of
the results from Belle and BABAR, current measurements yield a value for
$\alpha$ of about $90^\circ$ with an uncertainty of $\sim$7$^\circ$.  The decay
$\Bz\to\rho^\pm\pi^\mp$ can also be used to constrain $\alpha$ but this is
much less precise with present data.

\subsection{Measurements of \bma{$\gamma$}}
Several methods have been employed for measurements of the CKM angle
$\gamma$ that are theoretically clean.  All of these methods use decays
involving $D$ mesons such as $B\to DK$ and as such are beyond the scope
of this review.  The best of these measurements from Belle and BABAR find
$\gamma=76^\circ$ with uncertainties of $\sim$15$^\circ$.

Charmless hadronic $B$ decays are also sensitive to $\gamma$ and there
is a long history of suggested methods, all of which are imprecise with
current data.  The most successful methods have used global fits to many
of the branching fraction and \acp\ measurements discussed in this review.
A recent analysis of $PV$ decays \cite{ChiangPV} finds
$\gamma=(72\pm5)^\circ$, though this method is generally regarded to be
less robust due to flavor SU(3) breaking and other theoretical uncertainties.
Nevertheless, it is clear that measurements of the three angles sum to
$180^\circ$ within the experimental uncertainties of $\sim10^\circ$.

\subsection{Measurements of Penguin \bma{$\Bz$} Decays}
Possible physics beyond the Standard Model has been
intensively explored through measurements of time-dependent \CP\
asymmetries in the penguin $b\to s q\bar q$ decays such as $\Bz\to
(\phi,\omega,\pi^0,\etapr)K^0$.  In the SM, $S$ for these decays should
be nearly the same as the value measured for the $b\to c\bar c s$ decays such
as $\Bz\to J/\psi K^0$; there is a
small deviation {\it at most} ${\cal O}(0.1)$ \cite{London}. In order to
detect New Physics unambiguously in the
penguin $b\to s q\bar q$ modes, it is of great importance to understand SM
predictions for the difference $\Delta S_f\equiv -\eta_fS_f-S_{b\to c\bar c s}$
with $\eta_f=1$ ($-1$) for final \CP-even (odd) states.  The quantity $S_f$ has
been estimated in various QCD-based approaches; the results, together with
the measured values, are summarized in Table \ref{tab:DeltaS}. Since
$S_{b\to c\bar cs}=0.672\pm0.024$ \cite{HFAG}, it is clear that $\Delta S_f$
are predicted to be small and positive in most cases, while the experimental
central values of $\Delta S_f$ are negative except for the $K^+K^-K_S$ and
$K_SK_SK_S$ modes.  However the average of these measurements,
$0.64\pm0.04$ \cite{HFAG}, is less than one standard deviation below the
value from $b\to c\bar cs$ decays.

\begin{table}[!htbp]
\begin{center}
\caption{Mixing-induced \CP\ violation $S_f$ predicted in various approaches.
The QCDF results are taken from \cite{Beneke,CCS05,CCS3body,ChengSP}. There are
two solutions with some of SCET predictions.  The $K^+K^-K_S$ predictions
and measurements exclude the $\phi$ mass region.\newline}\label{tab:DeltaS}
\begin{tabular}{|l| c c c |c|}
\hline$-\eta_fS_f$
       &QCDF  
       &pQCD \cite{Li05,LiMish06}
       &SCET \cite{Zupan,Wang}
       &Expt \cite{BABARTD,BelleTD} 
       \\
       \hline
 $\phi K_S$
       & $0.75^{+0.00}_{-0.04}$
       & $0.71\pm0.01$
       & $0.69$
       & $0.44^{+0.17}_{-0.18}$
       \\
 $\omega K_S$
       & $0.85^{+0.03}_{-0.06}$
       & $0.84^{+0.03}_{-0.07}$
       & $\begin{array}{c} 0.50^{+0.05}_{-0.06}\\0.80\pm{0.02}\end{array}$
       & $0.45\pm{0.24}$
       \\
 $\rho^0K_S$
       & $0.64^{+0.03}_{-0.07}$
       & $0.50^{+0.10}_{-0.06}$
       & $\begin{array}{c}0.85^{+0.04}_{-0.05}\\ 0.56^{+0.02}_{-0.03}\end{array}$
       & $0.63_{-0.21}^{+0.17}$
       \\
 $\etapr K_S$
       & $0.74^{+0.00}_{-0.04}$
       &
       & $\begin{array}{c}0.706\pm0.008\\ 0.715\pm0.010\end{array}$
       & $0.60\pm0.07$
       \\
 $\eta K_S$
       & $0.79^{+0.02}_{-0.04}$
       &
       & $\begin{array}{c}0.69\pm0.16\\ 0.79\pm0.15\end{array}$
       &
       \\
 $\pi^0K_S$
       & $0.79^{+0.02}_{-0.04}$
       & $0.74^{+0.02}_{-0.03}$
       & $0.80\pm0.03$
       & $0.57\pm0.17$
       \\
 $f_0(980)K_S$
       & $0.731^{+0.001}_{-0.001}$
       &
       &  
       & $0.62^{+0.11}_{-0.13}$
       \\
  $K^+K^-K_S$
    & $0.728^{+0.009}_{-0.020}$ & &
    & $0.82\pm0.07$
    \\
  $K_SK_SK_S$
    & $0.719^{+0.009}_{-0.020}$ & &
    & $0.74\pm0.17$
    \\
  $K_S\pi^0\pi^0$
    & $0.729^{+0.009}_{-0.020}$ & &
    & $-0.52\pm0.41$
    \\
       \botrule
\end{tabular}
\end{center}
\end{table}

\section{CONCLUSIONS}

In this review, we have summarized branching fraction results for more
than 100 charmless $B$-meson decays.  Many of these decays have significant
experimental signals.  We also have shown results for $CP$-violating
asymmetry measurements for nearly 50 of these decays.  This represents
a truly impressive body of work, most of which have come from the Belle
and BABAR experiments in the last decade.

The global features of the branching fractions and CP asymmetries of these
charmless $B$ decays are generally well described by the QCD-motivated theories
such as QCDF, pQCD and SCET. The agreement between theory and experiment is
generally satisfactory. However, there remains some unsolved puzzles:
(i) $K\pi$ CP puzzle: it is naively expected that $B^+\to K^+\pi^0$ and
$\Bz\to K^+\pi^-$ have similar direct CP asymmetries, while they
differ by 5.3 $\sigma$ experimentally,
$\Delta A_{K\pi}=\acp(K^+\pi^0)-\acp(K^+\pi^-)=0.148\pm0.028$;
(ii) the abnormally large $B\to\etapr K$ rates: while the qualitative
picture of the enormously large rate of $B\to \etapr K$ over $B\to \eta K$ is
understood, a precise quantitative prediction is still lacking;
(iii) branching fractions of $\Bz\to\pi^0\pi^0,~\rho^0\pi^0$: theory usually
predicts too small rates for them and a reasonable one for $\Bz\to\rho^0\rho^0$;
(iv) the polarization puzzle in penguin-dominated $B\to VV$ decays: the
transverse polarization fraction is not as small as naively anticipated;
(v) mixing-induced $CP$ asymmetries: the predicted values of the effective
$\sin 2\beta$ for most of the $b\to ss\bar s$ induced decays are above the one
obtained from $B\to J/\psi K^0$, whereas experimentally the value of
$\sin2\beta$ from the bulk of the decay modes is systematically below that of
$B\to J/\psi K^0$; (vi) the angular correlation enigma in three-body baryonic $B$
decays: the short-distance picture of $b\to sg$ cannot explain or accommodate
the observed angular distributions in penguin-dominated decays such as
$B^+\to p\bar K^+,~p\bar\Lambda\pi^+$ and $\Bz\to \Lambda\bar\Lambda\Kz$.

The aforementioned puzzles pose a great challenge to the $B$-physics community
and their solutions need efforts from both theorists and experimentalists.
Either these enigmas can be resolved within the framework of the standard model
provided that the hadronic matrix elements are under fair control or new
physics effects already manifest themselves in some of the puzzles.

\section{ACKNOWLEDGMENTS}

We thank Rob Harr for help with the plots and Bill Ford and Hsiang-nan Li for
a careful reading of the manuscript. This work was supported in part by the
National Science Council of R.O.C. under grant NSC97-2112-M-001-004-MY3
and US Departmrnt of Energy under grant DE-FG02-04ER41290.


\end{document}